\documentclass[aps,prb,twocolumn]{revtex4-1}
\usepackage{graphicx}
\usepackage{amsmath}
\usepackage{amsfonts}
\usepackage{color}
\usepackage[colorlinks=true,linkcolor=blue,citecolor=blue,urlcolor=blue]{hyperref}

\newcommand{\zrte}{ZrTe$_5$}

\begin{document}

\title{Unraveling the topology of \zrte\ by changing temperature}

\author{Bartomeu Monserrat}
\email{bm418@cam.ac.uk}
\affiliation{Cavendish Laboratory, University of Cambridge, J. J. Thomson Avenue, Cambridge CB3 0HE, United Kingdom}

\author{Awadhesh Narayan}
\email{awadhesh@iisc.ac.in}
\affiliation{Solid State and Structural Chemistry Unit, Indian Institute of Science, Bangalore 560012, India}

\date{\today}

\begin{abstract}
We study the effects of temperature on the topological nature of \zrte, which sits near the phase boundary between strong and weak topological insulating orders. Using first-principles calculations, we show that the band gap and the topological indices of \zrte\ are extremely sensitive to thermal expansion, and that the electron-phonon interaction accounts for more than a third of the total temperature-dependent effects in \zrte. We find that the temperature dependence of the band gap has an opposite sign in the strong and weak topological insulator phases. Based on this insight, we propose a robust and unambiguous method to determine the topological nature of any given sample of \zrte: if the band gap decreases with temperature it is a strong topological insulator, and if it increases with temperature it is a weak topological insulator. An analogous strategy is expected to be generally applicable to other materials and to become particularly important in the vicinity of topological phase boundaries where other methods provide ambiguous results.
\end{abstract}

\maketitle

\section{Introduction} 

Topological insulators (TIs) and semimetals have become an important field of condensed matter over the last decade. TIs feature gapless surface states with spin-momentum locking that could be potentially useful for applications in spintronics and quantum computing \cite{hasan2010colloquium,qi2011topological}. More recently their semimetallic counterparts, which include Dirac and Weyl semimetals and various flavors of nodal semimetals, have gained widespread interest \cite{armitage2018weyl}. 

An intriguing addition to the family of topological matter was made by Weng and co-authors, who proposed monolayers of transition-metal pentatellurides \zrte\ and HfTe$_5$ as large gap quantum spin Hall insulators~\cite{weng2014transition}. This prediction has sparked intense experimental and theoretical activity on this material system, not only in the monolayer form but also in the bulk. Additionally, \zrte\ has proved a fertile ground for the discovery of a number of exciting properties: chiral magnetic effect~\cite{li2016chiral}, log-periodic quantum oscillations~\cite{wang2018discovery}, three-dimensional quantum Hall effect~\cite{tang2019three} and quantized thermoelectric Hall conductivity~\cite{zhang2019quantized}, to list just a few.

Despite this broad-ranging interest, the topological nature of \emph{bulk} \zrte\ has not been unambiguously identified and has led to considerable debate. Using angle-resolved photoemission spectroscopy (ARPES), Manzoni and co-workers have characterized \zrte\ as a strong TI \cite{manzoni2016evidence}. Other ARPES measurements show a weak TI phase \cite{moreschini2016arpes,zhang2017electronic,xiong2017three}, which is also supported by scanning tunneling microscopy \cite{li2016experimental,wu2016evidence} and Shubnikov-de-Haas oscillations \cite{lv2018shubnikov}. Other experiments based on magnetotransport \cite{zheng2016transport,liu2016zeeman,li2016chiral} and infrared spectroscopy \cite{chen2015infrared,chen2015magnetoinfrared} instead point to a Dirac semimetal phase. Yet other experiments using infrared spectroscopy observe a temperature-induced transition from a strong to a weak TI mediated by a Dirac semimetal \cite{xu2018temperature}. An associated debate is that surrounding the dimensionality of the proposed Dirac cone in \zrte\ and whether it is gapped or gapless~\cite{chen2015magnetoinfrared,yuan2016observation,chen2017spectroscopic,xiong2017three,pariari2017coexistence,jiang2017landau,zhang2019thermoelectric,martino2019two}. Many of these contradicting experimental results are supported by first-principles calculations, which raises the question of their reliability and predictive power.

In this contribution we study the effects of temperature on the topological nature of \zrte. We demonstrate that the band gap as well as the topological indices of this material are extremely sensitive to the volume and therefore to thermal expansion. Additionally, we calculate the change in the band gap of \zrte\ arising from electron-phonon coupling, which accounts for more than a third of the total temperature-dependent effects. We find that both volume and electron-phonon coupling corrections to the band gap have an opposite sign in the strong and weak TI regimes. Based on our results, we propose that monitoring the temperature dependence of the band gap provides an unambiguous way to determine the topological nature of any given sample of \zrte, which may be more generally applicable to materials close to topological phase boundaries.

\begin{figure*}
\includegraphics[scale=0.4]{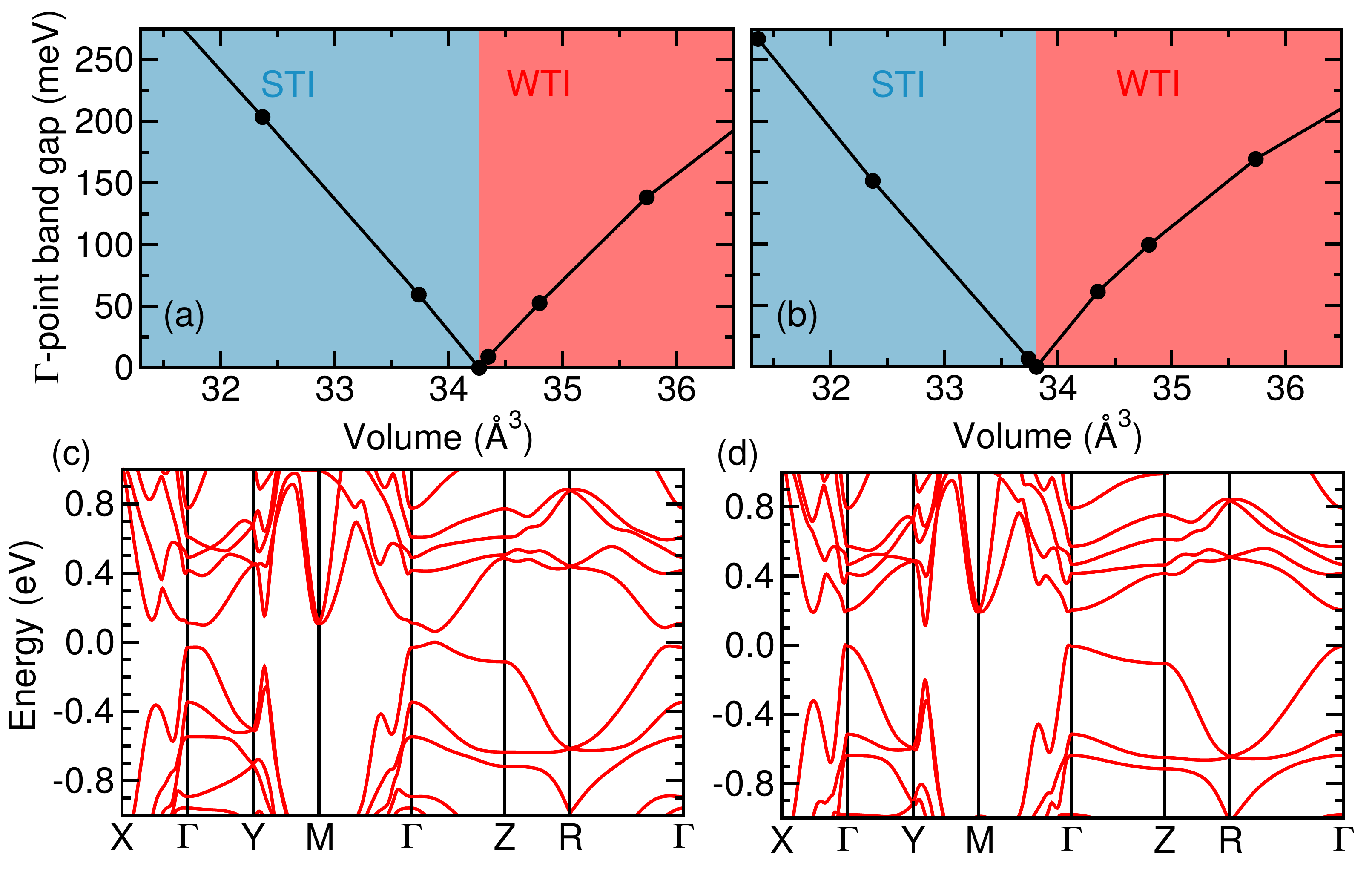}
  \caption{\textbf{Band gaps and band structures with volume.} Band gap at the $\Gamma$ point with increasing volume using (a) PBE and (b) HSE functionals. The shaded blue and red regions correspond to a strong topological insulator (STI) and a weak topological insulator (WTI) respectively. For both functionals the band gap first decreases with increasing volume, vanishing at the critical value marking a topological phase transition. The gap then reopens with further increase in volume beyond the critical value. Representative band structures in (c) strong and (d) weak TI regimes. }  \label{bands_gap}
\end{figure*}

\section{Methods} 

We use first-principles methods based on density functional theory (DFT) \cite{Hohenberg_Kohn,Kohn_Sham} as implemented in the {\sc vasp} code \cite{vasp1,vasp2,vasp3,vasp4} within the projector augmented-wave formalism \cite{paw_original,paw_us_relation}. Based on convergence tests, we use an energy cut-off of $500$\,eV and a Brillouin zone (BZ) Monkhorst-Pack \cite{MP_grid} $\mathbf{k}$-point grid of size $8\times8\times4$ for primitive cell calculations, and commensurate grids for supercell calculations. We describe the exchange-correlation functional with the generalized gradient approximation of Perdew-Burke-Ernzerhof (PBE) \cite{PBE} and also with the hybrid Heyd-Scuseria-Ernzerhof (HSE) functional \cite{hse06_functional,hse06_functional_erratum}. The spin-orbit interaction is included as a perturbation to the scalar relativistic Hamiltonian, the so-called second variational method \cite{soc_second_variation_method}.

Based on our first-principles results we construct a Wannier function based tight-binding model using the {\sc wannier90} code~\cite{mostofi2008wannier90}. We include the Zr $4d$ and Te $5p$ orbitals in the model. We then calculated the topological invariants for different structures using the {\sc WannierTools} package~\cite{wu2018wanniertools}.

The vibrational calculations are performed using the finite displacement method \cite{phonon_finite_displacement} in conjunction with nondiagonal supercells \cite{non_diagonal}. The matrix of force constants is explicitly calculated on a coarse $4\times4\times4$ $\mathbf{q}$-point grid, and then the dynamical matrix is obtained on a finer grid to calculate phonon dispersions. We note that the phonon calculations are very sensitive to the electronic BZ sampling grid. We perform the electron-phonon coupling calculations on a grid of $2\times2\times1$ $\mathbf{q}$-points.

\section{Results and Discussion} 

Temperature impacts the electronic structure and topology of materials through two main contributions, thermal expansion and electron-phonon coupling \cite{elph_topological_prl,elph_topological_prb,elph_topological_jhi,monserrat_ti_temp,antonius_ti_temp,wang_typeII_elph,monserrat2017temperature,moller_typeII_elph}.  Using our first-principles approach we are able to disentangle these two effects and look at their individual contributions. We note that electronic temperature can also play a role, but only very close to the transition itself \cite{topo_electronic_temp} and therefore we neglect this contribution in our calculations.

We find that both the band gap and the band topology of \zrte\ are extremely sensitive to the volume of the unit cell, and therefore to thermal expansion. Three-dimensional TIs are characterized by four indices $(\nu_{0};\nu_{1}\nu_{2}\nu_{3})$~\cite{hasan2010colloquium}. Based on the value of $\nu_{0}$, they can be classified into strong ($\nu_{0}=1$) and weak ($\nu_{0}=0$) TIs. At the relaxed PBE volume we find \zrte\ to be a weak TI with the topological indices $(0;110)$. On the other hand, for the experimentally-reported volumes at $10$\,K and $300$\,K \cite{zrte5_structure_Fjellvag_1986}, we find a strong TI phase in both cases. The band gap at the BZ centre ($\Gamma$) with increasing volume using the PBE functional is shown in Fig.\,\ref{bands_gap}(a). We find that, starting from small volumes, the band gap decreases with increasing volume in the strong TI phase. The gap vanishes at a critical volume, at which \zrte\ is a Dirac semimetal. The reopening of the band gap on further increase of the volume signals a topological phase transition into a weak TI. In the weak TI phase the band gap increases with increasing volume, a trend opposite to the case of the strong topological insulator phase. We also calculate the band structure and band topology using the HSE functional, which is often more reliable to estimate band gaps. The $\Gamma$-point band gap as a function of volume obtained with HSE is presented in Fig.\,\ref{bands_gap}(b). We find a similar behaviour of the band gap as a function of volume as in the case of PBE. While the critical value of volume at which the topological phase transition occurs is different by $\sim 1.3$\% using PBE and HSE, the trend of decreasing gap with increasing volume in the strong TI phase and the opposite behaviour in the weak TI regime are robust to the choice of theory.

We note that our results are consistent with recent theoretical studies discussing the effect of volume on the band structure of \zrte~\cite{weng2014transition,fan2017transition}. However, to fully capture temperature dependent material properties, the electron-phonon interaction also needs to be considered. As we will show next, these electron-phonon coupling effects are substantial and including them is essential for a complete understanding of the temperature dependence of the topological nature of \zrte.

\begin{figure}
\includegraphics[scale=0.45]{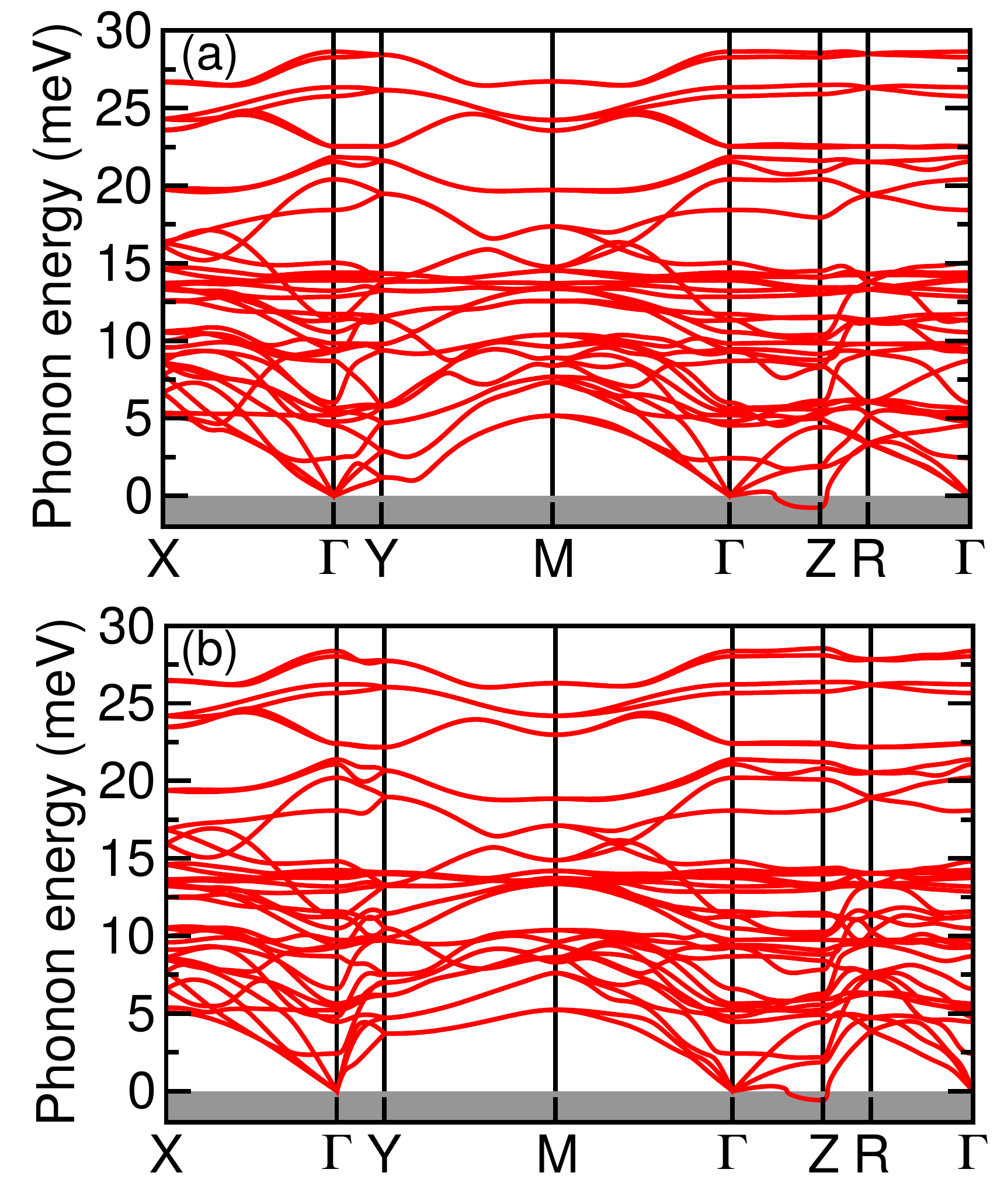}
  \caption{\textbf{Phonon spectra.} Phonon dispersions (a) without spin-orbit coupling and (b) with spin-orbit coupling along the high symmetry directions. Note the presence of an imaginary phonon frequency at Z in both cases.}  \label{phonon_bands}
\end{figure}

\begin{figure}
\includegraphics[scale=0.4]{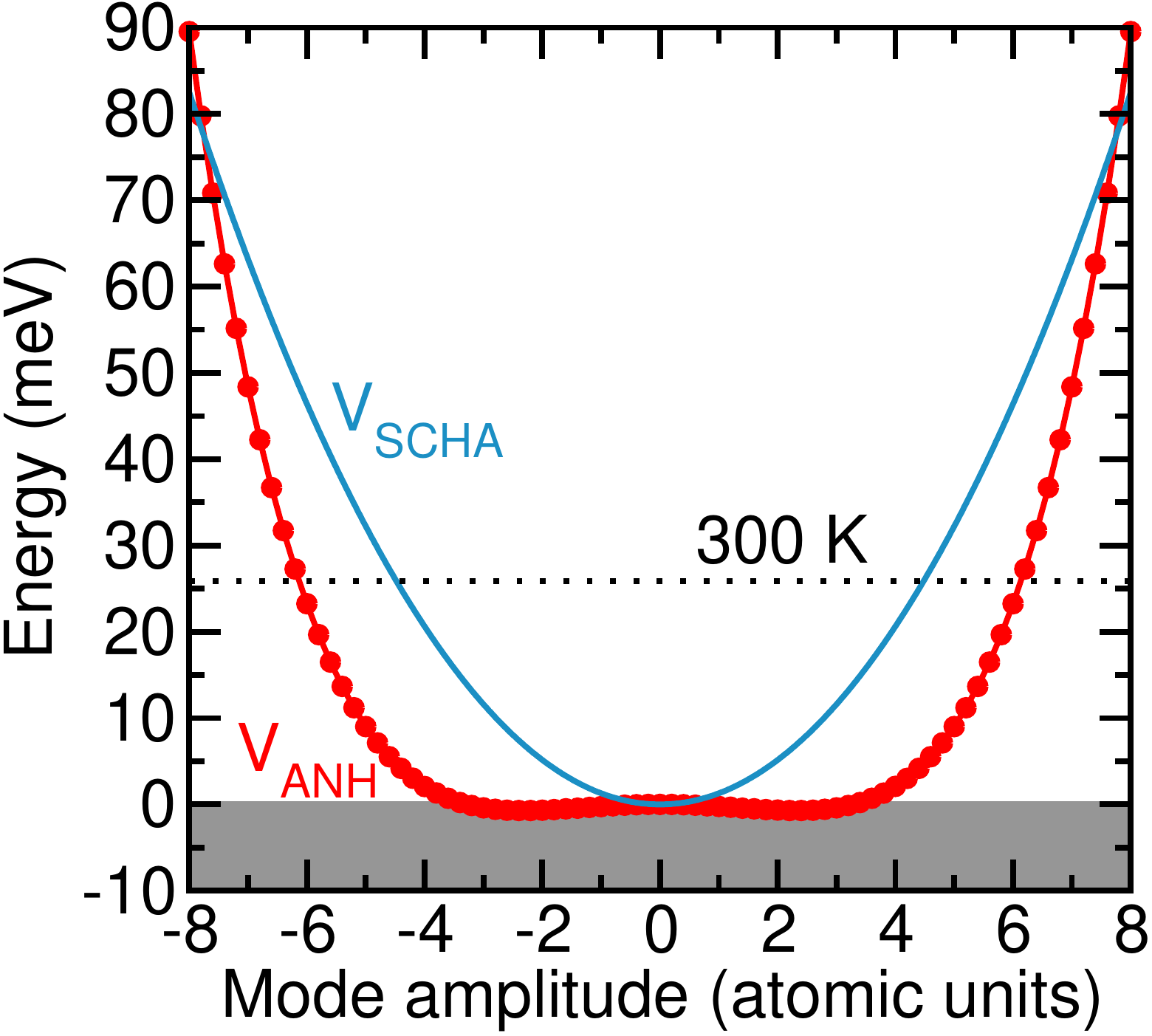}
  \caption{\textbf{Harmonic versus anharmonic approximations.} Energy as a function of the amplitude of the imaginary phonon mode. The red circles denote the anharmonic energy profile, which shows a shallow double well. The blue line is the self-consistent harmonic approximation, exhibiting a single minimum at zero amplitude. The black dashed line denotes the energy at 300 K.}  \label{scha}
\end{figure}

The phonon dispersion of \zrte\ is shown in Fig.\,\ref{phonon_bands} along a high symmetry path in the BZ obtained with {\sc seek-path} \cite{seek_path}. The top diagram shows the results from calculations performed without the inclusion of spin-orbit coupling, while the bottom diagram shows the corresponding results including spin-orbit coupling. The dispersions exhibit some differences, most prominently in the low-energy region along the path $\Gamma$--Y--M, where the phonons calculated including spin-orbit coupling have higher energies than those calculated without. Importantly, both dispersion curves exhibit an imaginary phonon mode at the Z point, which indicates that the $Cmcm$ structure of \zrte\ is not dynamically stable within the harmonic approximation, and that a lower energy structure of \zrte\ exists. Indeed, by distorting the $Cmcm$ structure along the imaginary phonon mode at Z and relaxing to the local minimum we obtain a structure of $Pnnm$ symmetry which is $0.7$\,meV per formula unit lower in energy than the initial structure. This result is at odds with experiments, which invariably report the higher symmetry $Cmcm$ structure for \zrte\ \cite{zrte5_structure_furuseth_1973,zrte5_structure_Sodeck_1979,zrte5_structure_Sambongi_1985,zrte5_structure_Fjellvag_1986}.

\begin{table*}[ht]
  \setlength{\tabcolsep}{6pt} 
\caption{\textbf{Band gap of \zrte\ at the $\Gamma$ point.} $E^{\mathrm{gap}}_{\mathrm{PBE}}$ and $E^{\mathrm{gap}}_{\mathrm{HSE}}$ are the static gaps using PBE and HSE functionals, respectively, and the correction due to electron-phonon coupling, calculated with the PBE functional, is denoted by $\Delta E^{\mathrm{gap}}_{\mathrm{el-ph}}$.} 
\label{gap_correction}
\begin{ruledtabular}
\begin{tabular}{lllccc}
Temperature & Volume & Topological indices & $E^{\mathrm{gap}}_{\mathrm{PBE}}$ (meV) & $E^{\mathrm{gap}}_{\mathrm{HSE}}$ (meV) & $\Delta E^{\mathrm{gap}}_{\mathrm{el-ph}}$ (meV) \\ 
\hline
$10$\,K  & Experiment at $10$\,K \cite{zrte5_structure_Fjellvag_1986}  & STI $(1;110)$ & $173$ & $122$ & $-16\pm1$ \\
$300$\,K & Experiment at $300$\,K \cite{zrte5_structure_Fjellvag_1986} & STI $(1;110)$ & $141$ & $91$  & $-33\pm2$ \\
$10$\,K  & PBE                            & WTI $(0;110)$ & $210$ & $224$ & $+3\pm1$ \\
$300$\,K & PBE                            & WTI $(0;110)$ & $210$ & $224$ & $+11\pm2$ \\
\end{tabular}
\end{ruledtabular}
\end{table*}

Fig.\,\ref{scha} shows the potential energy surface $V_{\mathrm{ANH}}$ of the $Cmcm$ structure of \zrte\ along the imaginary phonon mode at Z. This diagram shows that a distortion along this mode leads to an anharmonic double well potential. However, the energy scale of this anharmonic distortion is small compared to the thermal energy scale, which is $25.9$\,meV at $300$\,K (dotted black line in Fig.\,\ref{scha}). This suggests that anharmonic lattice dynamics might stabilize the experimentally observed $Cmcm$ structure of \zrte. To explore this possibility, we limit our analysis to the anharmonic mode at Z and use the self-consistent harmonic approximation (SCHA) \cite{scha}, which determines the effective harmonic free energy at some fixed temperature by including anharmonic terms in the total energy static potential. Consider a quartic anharmonic potential:
\begin{equation}
V_{\mathrm{ANH}}=\frac{1}{2}\omega^2u^2+\frac{1}{4}\lambda u^4, \label{eq:quartic}
\end{equation}
where $\omega^2<0$ for the unstable mode at Z. The self-consistent harmonic potential at inverse temperature $\beta$ is then $V_{\mathrm{SCHA}}=\frac{1}{2}\Omega^2u^2$, where \cite{scha_quartic}: 
\begin{equation}
\Omega^2=\omega^2+\frac{3\lambda}{2\Omega}\coth\left(\frac{\beta\Omega}{2}\right). \label{eq:scha}
\end{equation}
We fit Eq.\,(\ref{eq:quartic}) to the anharmonic potential energy surface shown in Fig.\,\ref{scha} to determine $\omega$ and $\lambda$. We then solve Eq.\,(\ref{eq:scha}) using a Newton-Raphson method, and the resulting potential $V_{\mathrm{SCHA}}$ at $300$\,K is depicted in Fig.\,\ref{scha}. This potential can be thought of as the free energy potential at $300$\,K, which exhibits a single well and therefore demonstrates that anharmonic thermal motion dynamically stabilises the experimentally observed $Cmcm$ structure of \zrte. Repeating this analysis at $10$\,K also leads to a single well free energy potential, indicating that even quantum zero-point motion is sufficient to stabilize the $Cmcm$ structure of \zrte.

We calculate the change in the band gap of \zrte\ at inverse temperature $\beta$ arising from electron-phonon coupling by evaluating the expectation value of electronic eigenvalues $\epsilon_{\mathbf{k}n}$, labeled by momentum $\mathbf{k}$ and band index $n$, with respect to the vibrational density:
\begin{equation}
\epsilon_{\mathbf{k}n}(\beta)=\int\prod_{\mathbf{q}\nu}\mathrm{d}u_{\mathbf{q}\nu}\epsilon_{\mathbf{k}n}(\{u_{\mathbf{q}\nu}\})|\phi_{\mathbf{q}\nu}(u_{\mathbf{q}\nu};\beta)|^2. \label{eq:elph}
\end{equation}
Phonon modes are labeled by their momentum $\mathbf{q}$ and branch index $\nu$, and the vibrational density is $|\phi_{\mathbf{q}\nu}(u_{\mathbf{q}\nu};\beta)|^2=[2\pi\sigma_{\mathbf{q}\nu}^2(\beta)]^{-1/2}\exp[-u_{\mathbf{q}\nu}^2/2\sigma_{\mathbf{q}\nu}^2(\beta)]$ with $\sigma_{\mathbf{q}\nu}^2(\beta)=(1/2\omega_{\mathbf{q}\nu})\coth(\beta\omega_{\mathbf{q}\nu}/2)$. In this expression, we treat the anharmonic phonon mode at Z within the self-consistent harmonic approximation, while all other phonon modes are treated within the harmonic approximation. We evaluate Eq.\,(\ref{eq:elph}) using stochastic integration accelerated with thermal lines \cite{thermal_lines} at $10$ and $300$\,K. We perform calculations for \zrte\ structures corresponding to the experimental volumes at $10$ and $300$\,K, and also for a fully relaxed structure using the PBE functional. The first two structures correspond to the regime in which \zrte\ is a strong TI, while the latter to the regime in which it is a weak TI. 

Table\,\ref{gap_correction} shows the correction to the band gap at the $\Gamma$ point of \zrte\ calculated from electron-phonon coupling. In the strong TI regime the band gap correction is negative, while in the weak TI regime the band correction is positive. The sign change in these corrections across the strong-to-weak topological transition arises because of the associated band inversion at the $\Gamma$ point. We also note that increasing temperature enhances the magnitude of the correction in both regimes. Comparing with other materials exhibiting topological order, the band gap change induced by electron-phonon coupling in \zrte\ has a similar magnitude to that of BiTeI \cite{monserrat2017temperature} and $\beta$-PbO$_2$ \cite{pbo2_temperature_topoloy}, while it is weaker than that in the Bi$_2$Se$_3$ family of topological insulators \cite{monserrat_ti_temp} and that in the alloy BiTl(S$_{1-\delta}$Se$_{\delta}$)$_2$ \cite{antonius_ti_temp}.

A full picture of the temperature dependence of the band gap of \zrte\ emerges from combining the thermal expansion and electron-phonon coupling results. In the strong TI regime, both thermal expansion and electron-phonon coupling contribute to decreasing the $\Gamma$-point band gap with increasing temperature, pushing \zrte\ towards the Dirac semimetal phase. In the weak TI regime, both thermal expansion and electron-phonon coupling contribute to increasing the $\Gamma$-point band gap with increasing temperature, moving \zrte\ away from the Dirac semimetal phase. We also note that increasing temperature leads to the closure of the band gap of the material, with the formation of a hole pocket at the $\Gamma$ point and an associated electron pocket elsewhere in the BZ. This temperature-dependent shift of the electronic bands is consistent with experimental reports~\cite{manzoni2015ultrafast} and has been connected to the resistivity anomaly in ZrTe$_5$~\cite{skelton1982giant,li2016chiral,zheng2016transport,liu2016zeeman,nair2018thermodynamic}.

It has proved challenging to determine the topological nature of \zrte. Experimentally, strong TI, weak TI, and Dirac semimetal phases have been proposed. It is possible that these reports are not inconsistent, with the differences arising for example from different growth conditions (see, for instance, supplementary material in Ref.\,\cite{wang2018discovery}). Many of these experiments rely on support from first-principles calculations to extract the topological nature of \zrte, which seems to suggest that these calculations can be used to support any desired outcome. As we have shown, the calculated topological order depends critically on the level of theory and on the volume of the structural models used. As a consequence, absolute comparisons are uncertain due to the many sensitive variables, and this might explain the contradicting reports found in the literature. To resolve this, we propose a reliable method to determine the topology of \zrte\ which does not rely on uncertain absolute values but instead on robust changes. Our results suggest an unambiguous way of determining the topological nature of any given sample of \zrte: if the band gap decreases with temperature it is a strong TI, and if the band gap increases with temperature it is a weak TI. A similar observation was made for BiTeI, which also shows an opposite sign in the temperature dependence of its band gap on either side of a pressure-induced topological phase transition~\cite{monserrat2017temperature}. We therefore propose that using temperature dependent changes in band gaps is a general and robust way to identify topological phases, specially in systems in the vicinity of topological phase boundaries where different experimental conditions might lead to distinct conclusions.

Our proposed model is consistent with available temperature-dependent experimental data. Zhang and co-workers performed temperature dependent angle-resolved photoemission measurements of the band gap of \zrte\ around the $\Gamma$-point \cite{zhang2017electronic}. They observe an increase of the gap with increasing temperature, which in our proposed model would imply that their sample is a weak TI. Indeed, the lack of topological surface states in the in-plane surface in their experiments is attributed to a weak TI bulk order \cite{zhang2017electronic}. Xu and co-workers performed a temperature-dependent infrared spectroscopy study of \zrte\ revealing a topological phase transition from a strong TI at low temperatures to a weak TI at high temperatures \cite{xu2018temperature}. Their observations can be consistently explained by our calculations incorporating both volume and electron-phonon coupling effects. Furthermore, they observe that the band gap decrease with temperature in the strong TI regime is stronger than the band gap increase with temperature in the weak TI regime. In our calculations, the volume-induced changes in the band gap are similar in both regimes and therefore cannot capture this experimental observation. Importantly, the inclusion of electron-phonon coupling in the model does explain the experimental observation, as the electron-phonon coupling contribution is significantly weaker in the weak TI regime (see Table\,\ref{gap_correction}).

\section{Summary} 

We have shown that the band topology of \zrte\ is extremely sensitive to volume and consequently to thermal expansion. We have also shown that the change in the band gap of \zrte\ due to electron-phonon coupling accounts for more than a third of the total temperature-dependent effects in this material. Overall, both thermal expansion and electron-phonon coupling corrections to the band gap have an opposite sign in the strong and weak TI phases of \zrte, explaining the available temperature-dependent experimental data. Based on these insights, we proposed a robust and unambiguous way to determine the topological nature of any given sample of \zrte\ by monitoring the temperature dependence of the band gap. This approach may be generally applicable to materials in the vicinity of topological phase boundaries, where other methods fail, as exemplified by the experimental controversies surrounding \zrte.

\section*{Acknowledgments} 

B.M. acknowledges financial support from the Winton Programme for the Physics of Sustainability, and from Robinson College, Cambridge, and the Cambridge Philosophical Society for a Henslow Research Fellowship. A.N. acknowledges support from the start-up grant at the Indian Institute of Science. Part of the calculations were performed using resources provided by the Cambridge Tier-2 system operated by the University of Cambridge Research Computing Service (http://www.hpc.cam.ac.uk) funded by EPSRC Tier-2 capital grant EP/P020259/1.


\begin{thebibliography}{65}%
\makeatletter
\providecommand \@ifxundefined [1]{%
 \@ifx{#1\undefined}
}%
\providecommand \@ifnum [1]{%
 \ifnum #1\expandafter \@firstoftwo
 \else \expandafter \@secondoftwo
 \fi
}%
\providecommand \@ifx [1]{%
 \ifx #1\expandafter \@firstoftwo
 \else \expandafter \@secondoftwo
 \fi
}%
\providecommand \natexlab [1]{#1}%
\providecommand \enquote  [1]{``#1''}%
\providecommand \bibnamefont  [1]{#1}%
\providecommand \bibfnamefont [1]{#1}%
\providecommand \citenamefont [1]{#1}%
\providecommand \href@noop [0]{\@secondoftwo}%
\providecommand \href [0]{\begingroup \@sanitize@url \@href}%
\providecommand \@href[1]{\@@startlink{#1}\@@href}%
\providecommand \@@href[1]{\endgroup#1\@@endlink}%
\providecommand \@sanitize@url [0]{\catcode `\\12\catcode `\$12\catcode
  `\&12\catcode `\#12\catcode `\^12\catcode `\_12\catcode `\%12\relax}%
\providecommand \@@startlink[1]{}%
\providecommand \@@endlink[0]{}%
\providecommand \url  [0]{\begingroup\@sanitize@url \@url }%
\providecommand \@url [1]{\endgroup\@href {#1}{\urlprefix }}%
\providecommand \urlprefix  [0]{URL }%
\providecommand \Eprint [0]{\href }%
\providecommand \doibase [0]{http://dx.doi.org/}%
\providecommand \selectlanguage [0]{\@gobble}%
\providecommand \bibinfo  [0]{\@secondoftwo}%
\providecommand \bibfield  [0]{\@secondoftwo}%
\providecommand \translation [1]{[#1]}%
\providecommand \BibitemOpen [0]{}%
\providecommand \bibitemStop [0]{}%
\providecommand \bibitemNoStop [0]{.\EOS\space}%
\providecommand \EOS [0]{\spacefactor3000\relax}%
\providecommand \BibitemShut  [1]{\csname bibitem#1\endcsname}%
\let\auto@bib@innerbib\@empty
\bibitem [{\citenamefont {Hasan}\ and\ \citenamefont
  {Kane}(2010)}]{hasan2010colloquium}%
  \BibitemOpen
  \bibfield  {author} {\bibinfo {author} {\bibfnamefont {M.~Z.}\ \bibnamefont
  {Hasan}}\ and\ \bibinfo {author} {\bibfnamefont {C.~L.}\ \bibnamefont
  {Kane}},\ }\href {https://link.aps.org/doi/10.1103/RevModPhys.82.3045}
  {\bibfield  {journal} {\bibinfo  {journal} {Rev. Mod. Phys.}\ }\textbf
  {\bibinfo {volume} {82}},\ \bibinfo {pages} {3045} (\bibinfo {year}
  {2010})}\BibitemShut {NoStop}%
\bibitem [{\citenamefont {Qi}\ and\ \citenamefont
  {Zhang}(2011)}]{qi2011topological}%
  \BibitemOpen
  \bibfield  {author} {\bibinfo {author} {\bibfnamefont {X.-L.}\ \bibnamefont
  {Qi}}\ and\ \bibinfo {author} {\bibfnamefont {S.-C.}\ \bibnamefont {Zhang}},\
  }\href {https://link.aps.org/doi/10.1103/RevModPhys.83.1057} {\bibfield
  {journal} {\bibinfo  {journal} {Rev. Mod. Phys.}\ }\textbf {\bibinfo {volume}
  {83}},\ \bibinfo {pages} {1057} (\bibinfo {year} {2011})}\BibitemShut
  {NoStop}%
\bibitem [{\citenamefont {Armitage}\ \emph {et~al.}(2018)\citenamefont
  {Armitage}, \citenamefont {Mele},\ and\ \citenamefont
  {Vishwanath}}]{armitage2018weyl}%
  \BibitemOpen
  \bibfield  {author} {\bibinfo {author} {\bibfnamefont {N.~P.}\ \bibnamefont
  {Armitage}}, \bibinfo {author} {\bibfnamefont {E.~J.}\ \bibnamefont {Mele}},
  \ and\ \bibinfo {author} {\bibfnamefont {A.}~\bibnamefont {Vishwanath}},\
  }\href {https://link.aps.org/doi/10.1103/RevModPhys.90.015001} {\bibfield
  {journal} {\bibinfo  {journal} {Rev. Mod. Phys.}\ }\textbf {\bibinfo {volume}
  {90}},\ \bibinfo {pages} {015001} (\bibinfo {year} {2018})}\BibitemShut
  {NoStop}%
\bibitem [{\citenamefont {Weng}\ \emph {et~al.}(2014)\citenamefont {Weng},
  \citenamefont {Dai},\ and\ \citenamefont {Fang}}]{weng2014transition}%
  \BibitemOpen
  \bibfield  {author} {\bibinfo {author} {\bibfnamefont {H.}~\bibnamefont
  {Weng}}, \bibinfo {author} {\bibfnamefont {X.}~\bibnamefont {Dai}}, \ and\
  \bibinfo {author} {\bibfnamefont {Z.}~\bibnamefont {Fang}},\ }\href
  {https://link.aps.org/doi/10.1103/PhysRevX.4.011002} {\bibfield  {journal}
  {\bibinfo  {journal} {Phys. Rev. X}\ }\textbf {\bibinfo {volume} {4}},\
  \bibinfo {pages} {011002} (\bibinfo {year} {2014})}\BibitemShut {NoStop}%
\bibitem [{\citenamefont {Li}\ \emph {et~al.}(2016{\natexlab{a}})\citenamefont
  {Li}, \citenamefont {Kharzeev}, \citenamefont {Zhang}, \citenamefont {Huang},
  \citenamefont {Pletikosi{\'c}}, \citenamefont {Fedorov}, \citenamefont
  {Zhong}, \citenamefont {Schneeloch}, \citenamefont {Gu},\ and\ \citenamefont
  {Valla}}]{li2016chiral}%
  \BibitemOpen
  \bibfield  {author} {\bibinfo {author} {\bibfnamefont {Q.}~\bibnamefont
  {Li}}, \bibinfo {author} {\bibfnamefont {D.~E.}\ \bibnamefont {Kharzeev}},
  \bibinfo {author} {\bibfnamefont {C.}~\bibnamefont {Zhang}}, \bibinfo
  {author} {\bibfnamefont {Y.}~\bibnamefont {Huang}}, \bibinfo {author}
  {\bibfnamefont {I.}~\bibnamefont {Pletikosi{\'c}}}, \bibinfo {author}
  {\bibfnamefont {A.~V.}\ \bibnamefont {Fedorov}}, \bibinfo {author}
  {\bibfnamefont {R.~D.}\ \bibnamefont {Zhong}}, \bibinfo {author}
  {\bibfnamefont {J.~A.}\ \bibnamefont {Schneeloch}}, \bibinfo {author}
  {\bibfnamefont {G.~D.}\ \bibnamefont {Gu}}, \ and\ \bibinfo {author}
  {\bibfnamefont {T.}~\bibnamefont {Valla}},\ }\href
  {https://doi.org/10.1038/nphys3648} {\bibfield  {journal} {\bibinfo
  {journal} {Nature Physics}\ }\textbf {\bibinfo {volume} {12}},\ \bibinfo
  {pages} {550} (\bibinfo {year} {2016}{\natexlab{a}})}\BibitemShut {NoStop}%
\bibitem [{\citenamefont {Wang}\ \emph {et~al.}(2018)\citenamefont {Wang},
  \citenamefont {Liu}, \citenamefont {Li}, \citenamefont {Liu}, \citenamefont
  {Wang}, \citenamefont {Liu}, \citenamefont {Dai}, \citenamefont {Wang},
  \citenamefont {Li}, \citenamefont {Yan} \emph {et~al.}}]{wang2018discovery}%
  \BibitemOpen
  \bibfield  {author} {\bibinfo {author} {\bibfnamefont {H.}~\bibnamefont
  {Wang}}, \bibinfo {author} {\bibfnamefont {H.}~\bibnamefont {Liu}}, \bibinfo
  {author} {\bibfnamefont {Y.}~\bibnamefont {Li}}, \bibinfo {author}
  {\bibfnamefont {Y.}~\bibnamefont {Liu}}, \bibinfo {author} {\bibfnamefont
  {J.}~\bibnamefont {Wang}}, \bibinfo {author} {\bibfnamefont {J.}~\bibnamefont
  {Liu}}, \bibinfo {author} {\bibfnamefont {J.-Y.}\ \bibnamefont {Dai}},
  \bibinfo {author} {\bibfnamefont {Y.}~\bibnamefont {Wang}}, \bibinfo {author}
  {\bibfnamefont {L.}~\bibnamefont {Li}}, \bibinfo {author} {\bibfnamefont
  {J.}~\bibnamefont {Yan}},  \emph {et~al.},\ }\href
  {https://advances.sciencemag.org/content/4/11/eaau5096} {\bibfield  {journal}
  {\bibinfo  {journal} {Sci. Adv.}\ }\textbf {\bibinfo {volume} {4}},\ \bibinfo
  {pages} {eaau5096} (\bibinfo {year} {2018})}\BibitemShut {NoStop}%
\bibitem [{\citenamefont {Tang}\ \emph {et~al.}(2019)\citenamefont {Tang},
  \citenamefont {Ren}, \citenamefont {Wang}, \citenamefont {Zhong},
  \citenamefont {Schneeloch}, \citenamefont {Yang}, \citenamefont {Yang},
  \citenamefont {Lee}, \citenamefont {Gu}, \citenamefont {Qiao},\ and\
  \citenamefont {Zhang}}]{tang2019three}%
  \BibitemOpen
  \bibfield  {author} {\bibinfo {author} {\bibfnamefont {F.}~\bibnamefont
  {Tang}}, \bibinfo {author} {\bibfnamefont {Y.}~\bibnamefont {Ren}}, \bibinfo
  {author} {\bibfnamefont {P.}~\bibnamefont {Wang}}, \bibinfo {author}
  {\bibfnamefont {R.}~\bibnamefont {Zhong}}, \bibinfo {author} {\bibfnamefont
  {J.}~\bibnamefont {Schneeloch}}, \bibinfo {author} {\bibfnamefont {S.~A.}\
  \bibnamefont {Yang}}, \bibinfo {author} {\bibfnamefont {K.}~\bibnamefont
  {Yang}}, \bibinfo {author} {\bibfnamefont {P.~A.}\ \bibnamefont {Lee}},
  \bibinfo {author} {\bibfnamefont {G.}~\bibnamefont {Gu}}, \bibinfo {author}
  {\bibfnamefont {Z.}~\bibnamefont {Qiao}}, \ and\ \bibinfo {author}
  {\bibfnamefont {L.}~\bibnamefont {Zhang}},\ }\href
  {https://doi.org/10.1038/s41586-019-1180-9} {\bibfield  {journal} {\bibinfo
  {journal} {Nature}\ }\textbf {\bibinfo {volume} {569}},\ \bibinfo {pages}
  {537} (\bibinfo {year} {2019})}\BibitemShut {NoStop}%
\bibitem [{\citenamefont {Zhang}\ \emph
  {et~al.}(2019{\natexlab{a}})\citenamefont {Zhang}, \citenamefont {Wang},
  \citenamefont {Skinner}, \citenamefont {Bi}, \citenamefont {Kozii},
  \citenamefont {Cho}, \citenamefont {Zhong}, \citenamefont {Schneeloch},
  \citenamefont {Yu}, \citenamefont {Gu}, \citenamefont {Fu}, \citenamefont
  {Wu},\ and\ \citenamefont {Zhang}}]{zhang2019quantized}%
  \BibitemOpen
  \bibfield  {author} {\bibinfo {author} {\bibfnamefont {W.}~\bibnamefont
  {Zhang}}, \bibinfo {author} {\bibfnamefont {P.}~\bibnamefont {Wang}},
  \bibinfo {author} {\bibfnamefont {B.}~\bibnamefont {Skinner}}, \bibinfo
  {author} {\bibfnamefont {R.}~\bibnamefont {Bi}}, \bibinfo {author}
  {\bibfnamefont {V.}~\bibnamefont {Kozii}}, \bibinfo {author} {\bibfnamefont
  {C.-W.}\ \bibnamefont {Cho}}, \bibinfo {author} {\bibfnamefont
  {R.}~\bibnamefont {Zhong}}, \bibinfo {author} {\bibfnamefont
  {J.}~\bibnamefont {Schneeloch}}, \bibinfo {author} {\bibfnamefont
  {D.}~\bibnamefont {Yu}}, \bibinfo {author} {\bibfnamefont {G.}~\bibnamefont
  {Gu}}, \bibinfo {author} {\bibfnamefont {L.}~\bibnamefont {Fu}}, \bibinfo
  {author} {\bibfnamefont {X.}~\bibnamefont {Wu}}, \ and\ \bibinfo {author}
  {\bibfnamefont {L.}~\bibnamefont {Zhang}},\ }\href
  {https://arxiv.org/abs/1904.02157} {\bibfield  {journal} {\bibinfo  {journal}
  {arXiv:1904.02157}\ } (\bibinfo {year} {2019}{\natexlab{a}})}\BibitemShut
  {NoStop}%
\bibitem [{\citenamefont {Manzoni}\ \emph {et~al.}(2016)\citenamefont
  {Manzoni}, \citenamefont {Gragnaniello}, \citenamefont {Aut\`es},
  \citenamefont {Kuhn}, \citenamefont {Sterzi}, \citenamefont {Cilento},
  \citenamefont {Zacchigna}, \citenamefont {Enenkel}, \citenamefont {Vobornik},
  \citenamefont {Barba}, \citenamefont {Bisti}, \citenamefont {Bugnon},
  \citenamefont {Magrez}, \citenamefont {Strocov}, \citenamefont {Berger},
  \citenamefont {Yazyev}, \citenamefont {Fonin}, \citenamefont {Parmigiani},\
  and\ \citenamefont {Crepaldi}}]{manzoni2016evidence}%
  \BibitemOpen
  \bibfield  {author} {\bibinfo {author} {\bibfnamefont {G.}~\bibnamefont
  {Manzoni}}, \bibinfo {author} {\bibfnamefont {L.}~\bibnamefont
  {Gragnaniello}}, \bibinfo {author} {\bibfnamefont {G.}~\bibnamefont
  {Aut\`es}}, \bibinfo {author} {\bibfnamefont {T.}~\bibnamefont {Kuhn}},
  \bibinfo {author} {\bibfnamefont {A.}~\bibnamefont {Sterzi}}, \bibinfo
  {author} {\bibfnamefont {F.}~\bibnamefont {Cilento}}, \bibinfo {author}
  {\bibfnamefont {M.}~\bibnamefont {Zacchigna}}, \bibinfo {author}
  {\bibfnamefont {V.}~\bibnamefont {Enenkel}}, \bibinfo {author} {\bibfnamefont
  {I.}~\bibnamefont {Vobornik}}, \bibinfo {author} {\bibfnamefont
  {L.}~\bibnamefont {Barba}}, \bibinfo {author} {\bibfnamefont
  {F.}~\bibnamefont {Bisti}}, \bibinfo {author} {\bibfnamefont
  {P.}~\bibnamefont {Bugnon}}, \bibinfo {author} {\bibfnamefont
  {A.}~\bibnamefont {Magrez}}, \bibinfo {author} {\bibfnamefont {V.~N.}\
  \bibnamefont {Strocov}}, \bibinfo {author} {\bibfnamefont {H.}~\bibnamefont
  {Berger}}, \bibinfo {author} {\bibfnamefont {O.~V.}\ \bibnamefont {Yazyev}},
  \bibinfo {author} {\bibfnamefont {M.}~\bibnamefont {Fonin}}, \bibinfo
  {author} {\bibfnamefont {F.}~\bibnamefont {Parmigiani}}, \ and\ \bibinfo
  {author} {\bibfnamefont {A.}~\bibnamefont {Crepaldi}},\ }\href
  {https://link.aps.org/doi/10.1103/PhysRevLett.117.237601} {\bibfield
  {journal} {\bibinfo  {journal} {Phys. Rev. Lett.}\ }\textbf {\bibinfo
  {volume} {117}},\ \bibinfo {pages} {237601} (\bibinfo {year}
  {2016})}\BibitemShut {NoStop}%
\bibitem [{\citenamefont {Moreschini}\ \emph {et~al.}(2016)\citenamefont
  {Moreschini}, \citenamefont {Johannsen}, \citenamefont {Berger},
  \citenamefont {Denlinger}, \citenamefont {Jozwiak}, \citenamefont
  {Rotenberg}, \citenamefont {Kim}, \citenamefont {Bostwick},\ and\
  \citenamefont {Grioni}}]{moreschini2016arpes}%
  \BibitemOpen
  \bibfield  {author} {\bibinfo {author} {\bibfnamefont {L.}~\bibnamefont
  {Moreschini}}, \bibinfo {author} {\bibfnamefont {J.~C.}\ \bibnamefont
  {Johannsen}}, \bibinfo {author} {\bibfnamefont {H.}~\bibnamefont {Berger}},
  \bibinfo {author} {\bibfnamefont {J.}~\bibnamefont {Denlinger}}, \bibinfo
  {author} {\bibfnamefont {C.}~\bibnamefont {Jozwiak}}, \bibinfo {author}
  {\bibfnamefont {E.}~\bibnamefont {Rotenberg}}, \bibinfo {author}
  {\bibfnamefont {K.~S.}\ \bibnamefont {Kim}}, \bibinfo {author} {\bibfnamefont
  {A.}~\bibnamefont {Bostwick}}, \ and\ \bibinfo {author} {\bibfnamefont
  {M.}~\bibnamefont {Grioni}},\ }\href
  {https://link.aps.org/doi/10.1103/PhysRevB.94.081101} {\bibfield  {journal}
  {\bibinfo  {journal} {Phys. Rev. B}\ }\textbf {\bibinfo {volume} {94}},\
  \bibinfo {pages} {081101} (\bibinfo {year} {2016})}\BibitemShut {NoStop}%
\bibitem [{\citenamefont {Zhang}\ \emph {et~al.}(2017)\citenamefont {Zhang},
  \citenamefont {Wang}, \citenamefont {Yu}, \citenamefont {Liu}, \citenamefont
  {Liang}, \citenamefont {Huang}, \citenamefont {Nie}, \citenamefont {Sun},
  \citenamefont {Zhang}, \citenamefont {Shen}, \citenamefont {Liu},
  \citenamefont {Weng}, \citenamefont {Zhao}, \citenamefont {Chen},
  \citenamefont {Jia}, \citenamefont {Hu}, \citenamefont {Ding}, \citenamefont
  {Zhao}, \citenamefont {Gao}, \citenamefont {Li}, \citenamefont {He},
  \citenamefont {Zhao}, \citenamefont {Zhang}, \citenamefont {Zhang},
  \citenamefont {Yang}, \citenamefont {Wang}, \citenamefont {Peng},
  \citenamefont {Dai}, \citenamefont {Fang}, \citenamefont {Xu}, \citenamefont
  {Chen},\ and\ \citenamefont {Zhou}}]{zhang2017electronic}%
  \BibitemOpen
  \bibfield  {author} {\bibinfo {author} {\bibfnamefont {Y.}~\bibnamefont
  {Zhang}}, \bibinfo {author} {\bibfnamefont {C.}~\bibnamefont {Wang}},
  \bibinfo {author} {\bibfnamefont {L.}~\bibnamefont {Yu}}, \bibinfo {author}
  {\bibfnamefont {G.}~\bibnamefont {Liu}}, \bibinfo {author} {\bibfnamefont
  {A.}~\bibnamefont {Liang}}, \bibinfo {author} {\bibfnamefont
  {J.}~\bibnamefont {Huang}}, \bibinfo {author} {\bibfnamefont
  {S.}~\bibnamefont {Nie}}, \bibinfo {author} {\bibfnamefont {X.}~\bibnamefont
  {Sun}}, \bibinfo {author} {\bibfnamefont {Y.}~\bibnamefont {Zhang}}, \bibinfo
  {author} {\bibfnamefont {B.}~\bibnamefont {Shen}}, \bibinfo {author}
  {\bibfnamefont {J.}~\bibnamefont {Liu}}, \bibinfo {author} {\bibfnamefont
  {H.}~\bibnamefont {Weng}}, \bibinfo {author} {\bibfnamefont {L.}~\bibnamefont
  {Zhao}}, \bibinfo {author} {\bibfnamefont {G.}~\bibnamefont {Chen}}, \bibinfo
  {author} {\bibfnamefont {X.}~\bibnamefont {Jia}}, \bibinfo {author}
  {\bibfnamefont {C.}~\bibnamefont {Hu}}, \bibinfo {author} {\bibfnamefont
  {Y.}~\bibnamefont {Ding}}, \bibinfo {author} {\bibfnamefont {W.}~\bibnamefont
  {Zhao}}, \bibinfo {author} {\bibfnamefont {Q.}~\bibnamefont {Gao}}, \bibinfo
  {author} {\bibfnamefont {C.}~\bibnamefont {Li}}, \bibinfo {author}
  {\bibfnamefont {S.}~\bibnamefont {He}}, \bibinfo {author} {\bibfnamefont
  {L.}~\bibnamefont {Zhao}}, \bibinfo {author} {\bibfnamefont {F.}~\bibnamefont
  {Zhang}}, \bibinfo {author} {\bibfnamefont {S.}~\bibnamefont {Zhang}},
  \bibinfo {author} {\bibfnamefont {F.}~\bibnamefont {Yang}}, \bibinfo {author}
  {\bibfnamefont {Z.}~\bibnamefont {Wang}}, \bibinfo {author} {\bibfnamefont
  {Q.}~\bibnamefont {Peng}}, \bibinfo {author} {\bibfnamefont {X.}~\bibnamefont
  {Dai}}, \bibinfo {author} {\bibfnamefont {Z.}~\bibnamefont {Fang}}, \bibinfo
  {author} {\bibfnamefont {Z.}~\bibnamefont {Xu}}, \bibinfo {author}
  {\bibfnamefont {C.}~\bibnamefont {Chen}}, \ and\ \bibinfo {author}
  {\bibfnamefont {X.~J.}\ \bibnamefont {Zhou}},\ }\href
  {https://doi.org/10.1038/ncomms15512} {\bibfield  {journal} {\bibinfo
  {journal} {Nature Commun.}\ }\textbf {\bibinfo {volume} {8}},\ \bibinfo
  {pages} {15512} (\bibinfo {year} {2017})}\BibitemShut {NoStop}%
\bibitem [{\citenamefont {Xiong}\ \emph {et~al.}(2017)\citenamefont {Xiong},
  \citenamefont {Sobota}, \citenamefont {Yang}, \citenamefont {Soifer},
  \citenamefont {Gauthier}, \citenamefont {Lu}, \citenamefont {Lv},
  \citenamefont {Yao}, \citenamefont {Lu}, \citenamefont {Hashimoto},
  \citenamefont {Kirchmann}, \citenamefont {Chen},\ and\ \citenamefont
  {Shen}}]{xiong2017three}%
  \BibitemOpen
  \bibfield  {author} {\bibinfo {author} {\bibfnamefont {H.}~\bibnamefont
  {Xiong}}, \bibinfo {author} {\bibfnamefont {J.~A.}\ \bibnamefont {Sobota}},
  \bibinfo {author} {\bibfnamefont {S.-L.}\ \bibnamefont {Yang}}, \bibinfo
  {author} {\bibfnamefont {H.}~\bibnamefont {Soifer}}, \bibinfo {author}
  {\bibfnamefont {A.}~\bibnamefont {Gauthier}}, \bibinfo {author}
  {\bibfnamefont {M.-H.}\ \bibnamefont {Lu}}, \bibinfo {author} {\bibfnamefont
  {Y.-Y.}\ \bibnamefont {Lv}}, \bibinfo {author} {\bibfnamefont {S.-H.}\
  \bibnamefont {Yao}}, \bibinfo {author} {\bibfnamefont {D.}~\bibnamefont
  {Lu}}, \bibinfo {author} {\bibfnamefont {M.}~\bibnamefont {Hashimoto}},
  \bibinfo {author} {\bibfnamefont {P.~S.}\ \bibnamefont {Kirchmann}}, \bibinfo
  {author} {\bibfnamefont {Y.-F.}\ \bibnamefont {Chen}}, \ and\ \bibinfo
  {author} {\bibfnamefont {Z.-X.}\ \bibnamefont {Shen}},\ }\href
  {https://link.aps.org/doi/10.1103/PhysRevB.95.195119} {\bibfield  {journal}
  {\bibinfo  {journal} {Physical Review B}\ }\textbf {\bibinfo {volume} {95}},\
  \bibinfo {pages} {195119} (\bibinfo {year} {2017})}\BibitemShut {NoStop}%
\bibitem [{\citenamefont {Li}\ \emph {et~al.}(2016{\natexlab{b}})\citenamefont
  {Li}, \citenamefont {Huang}, \citenamefont {Lv}, \citenamefont {Zhang},
  \citenamefont {Yang}, \citenamefont {Zhang}, \citenamefont {Chen},
  \citenamefont {Yao}, \citenamefont {Zhou}, \citenamefont {Lu}, \citenamefont
  {Sheng}, \citenamefont {Li}, \citenamefont {Jia}, \citenamefont {Xue},
  \citenamefont {Chen},\ and\ \citenamefont {Xing}}]{li2016experimental}%
  \BibitemOpen
  \bibfield  {author} {\bibinfo {author} {\bibfnamefont {X.-B.}\ \bibnamefont
  {Li}}, \bibinfo {author} {\bibfnamefont {W.-K.}\ \bibnamefont {Huang}},
  \bibinfo {author} {\bibfnamefont {Y.-Y.}\ \bibnamefont {Lv}}, \bibinfo
  {author} {\bibfnamefont {K.-W.}\ \bibnamefont {Zhang}}, \bibinfo {author}
  {\bibfnamefont {C.-L.}\ \bibnamefont {Yang}}, \bibinfo {author}
  {\bibfnamefont {B.-B.}\ \bibnamefont {Zhang}}, \bibinfo {author}
  {\bibfnamefont {Y.~B.}\ \bibnamefont {Chen}}, \bibinfo {author}
  {\bibfnamefont {S.-H.}\ \bibnamefont {Yao}}, \bibinfo {author} {\bibfnamefont
  {J.}~\bibnamefont {Zhou}}, \bibinfo {author} {\bibfnamefont {M.-H.}\
  \bibnamefont {Lu}}, \bibinfo {author} {\bibfnamefont {L.}~\bibnamefont
  {Sheng}}, \bibinfo {author} {\bibfnamefont {S.-C.}\ \bibnamefont {Li}},
  \bibinfo {author} {\bibfnamefont {J.-F.}\ \bibnamefont {Jia}}, \bibinfo
  {author} {\bibfnamefont {Q.-K.}\ \bibnamefont {Xue}}, \bibinfo {author}
  {\bibfnamefont {Y.-F.}\ \bibnamefont {Chen}}, \ and\ \bibinfo {author}
  {\bibfnamefont {D.-Y.}\ \bibnamefont {Xing}},\ }\href
  {https://link.aps.org/doi/10.1103/PhysRevLett.116.176803} {\bibfield
  {journal} {\bibinfo  {journal} {Phys. Rev. Lett.}\ }\textbf {\bibinfo
  {volume} {116}},\ \bibinfo {pages} {176803} (\bibinfo {year}
  {2016}{\natexlab{b}})}\BibitemShut {NoStop}%
\bibitem [{\citenamefont {Wu}\ \emph {et~al.}(2016)\citenamefont {Wu},
  \citenamefont {Ma}, \citenamefont {Nie}, \citenamefont {Zhao}, \citenamefont
  {Huang}, \citenamefont {Yin}, \citenamefont {Fu}, \citenamefont {Richard},
  \citenamefont {Chen}, \citenamefont {Fang}, \citenamefont {Dai},
  \citenamefont {Weng}, \citenamefont {Qian}, \citenamefont {Ding},\ and\
  \citenamefont {Pan}}]{wu2016evidence}%
  \BibitemOpen
  \bibfield  {author} {\bibinfo {author} {\bibfnamefont {R.}~\bibnamefont
  {Wu}}, \bibinfo {author} {\bibfnamefont {J.-Z.}\ \bibnamefont {Ma}}, \bibinfo
  {author} {\bibfnamefont {S.-M.}\ \bibnamefont {Nie}}, \bibinfo {author}
  {\bibfnamefont {L.-X.}\ \bibnamefont {Zhao}}, \bibinfo {author}
  {\bibfnamefont {X.}~\bibnamefont {Huang}}, \bibinfo {author} {\bibfnamefont
  {J.-X.}\ \bibnamefont {Yin}}, \bibinfo {author} {\bibfnamefont {B.-B.}\
  \bibnamefont {Fu}}, \bibinfo {author} {\bibfnamefont {P.}~\bibnamefont
  {Richard}}, \bibinfo {author} {\bibfnamefont {G.-F.}\ \bibnamefont {Chen}},
  \bibinfo {author} {\bibfnamefont {Z.}~\bibnamefont {Fang}}, \bibinfo {author}
  {\bibfnamefont {X.}~\bibnamefont {Dai}}, \bibinfo {author} {\bibfnamefont
  {H.-M.}\ \bibnamefont {Weng}}, \bibinfo {author} {\bibfnamefont
  {T.}~\bibnamefont {Qian}}, \bibinfo {author} {\bibfnamefont {H.}~\bibnamefont
  {Ding}}, \ and\ \bibinfo {author} {\bibfnamefont {S.~H.}\ \bibnamefont
  {Pan}},\ }\href {https://link.aps.org/doi/10.1103/PhysRevX.6.021017}
  {\bibfield  {journal} {\bibinfo  {journal} {Phys. Rev. X}\ }\textbf {\bibinfo
  {volume} {6}},\ \bibinfo {pages} {021017} (\bibinfo {year}
  {2016})}\BibitemShut {NoStop}%
\bibitem [{\citenamefont {Lv}\ \emph {et~al.}(2018)\citenamefont {Lv},
  \citenamefont {Zhang}, \citenamefont {Li}, \citenamefont {Zhang},
  \citenamefont {Li}, \citenamefont {Yao}, \citenamefont {Chen}, \citenamefont
  {Zhou}, \citenamefont {Zhang}, \citenamefont {Lu}, \citenamefont {Li},\ and\
  \citenamefont {Chen}}]{lv2018shubnikov}%
  \BibitemOpen
  \bibfield  {author} {\bibinfo {author} {\bibfnamefont {Y.-Y.}\ \bibnamefont
  {Lv}}, \bibinfo {author} {\bibfnamefont {B.-B.}\ \bibnamefont {Zhang}},
  \bibinfo {author} {\bibfnamefont {X.}~\bibnamefont {Li}}, \bibinfo {author}
  {\bibfnamefont {K.-W.}\ \bibnamefont {Zhang}}, \bibinfo {author}
  {\bibfnamefont {X.-B.}\ \bibnamefont {Li}}, \bibinfo {author} {\bibfnamefont
  {S.-H.}\ \bibnamefont {Yao}}, \bibinfo {author} {\bibfnamefont {Y.~B.}\
  \bibnamefont {Chen}}, \bibinfo {author} {\bibfnamefont {J.}~\bibnamefont
  {Zhou}}, \bibinfo {author} {\bibfnamefont {S.-T.}\ \bibnamefont {Zhang}},
  \bibinfo {author} {\bibfnamefont {M.-H.}\ \bibnamefont {Lu}}, \bibinfo
  {author} {\bibfnamefont {S.-C.}\ \bibnamefont {Li}}, \ and\ \bibinfo {author}
  {\bibfnamefont {Y.-F.}\ \bibnamefont {Chen}},\ }\href
  {https://link.aps.org/doi/10.1103/PhysRevB.97.115137} {\bibfield  {journal}
  {\bibinfo  {journal} {Phys. Rev. B}\ }\textbf {\bibinfo {volume} {97}},\
  \bibinfo {pages} {115137} (\bibinfo {year} {2018})}\BibitemShut {NoStop}%
\bibitem [{\citenamefont {Zheng}\ \emph {et~al.}(2016)\citenamefont {Zheng},
  \citenamefont {Lu}, \citenamefont {Zhu}, \citenamefont {Ning}, \citenamefont
  {Han}, \citenamefont {Zhang}, \citenamefont {Zhang}, \citenamefont {Xi},
  \citenamefont {Yang}, \citenamefont {Du}, \citenamefont {Yang}, \citenamefont
  {Zhang},\ and\ \citenamefont {Tian}}]{zheng2016transport}%
  \BibitemOpen
  \bibfield  {author} {\bibinfo {author} {\bibfnamefont {G.}~\bibnamefont
  {Zheng}}, \bibinfo {author} {\bibfnamefont {J.}~\bibnamefont {Lu}}, \bibinfo
  {author} {\bibfnamefont {X.}~\bibnamefont {Zhu}}, \bibinfo {author}
  {\bibfnamefont {W.}~\bibnamefont {Ning}}, \bibinfo {author} {\bibfnamefont
  {Y.}~\bibnamefont {Han}}, \bibinfo {author} {\bibfnamefont {H.}~\bibnamefont
  {Zhang}}, \bibinfo {author} {\bibfnamefont {J.}~\bibnamefont {Zhang}},
  \bibinfo {author} {\bibfnamefont {C.}~\bibnamefont {Xi}}, \bibinfo {author}
  {\bibfnamefont {J.}~\bibnamefont {Yang}}, \bibinfo {author} {\bibfnamefont
  {H.}~\bibnamefont {Du}}, \bibinfo {author} {\bibfnamefont {K.}~\bibnamefont
  {Yang}}, \bibinfo {author} {\bibfnamefont {Y.}~\bibnamefont {Zhang}}, \ and\
  \bibinfo {author} {\bibfnamefont {M.}~\bibnamefont {Tian}},\ }\href
  {https://link.aps.org/doi/10.1103/PhysRevB.93.115414} {\bibfield  {journal}
  {\bibinfo  {journal} {Phys. Rev. B}\ }\textbf {\bibinfo {volume} {93}},\
  \bibinfo {pages} {115414} (\bibinfo {year} {2016})}\BibitemShut {NoStop}%
\bibitem [{\citenamefont {Liu}\ \emph {et~al.}(2016)\citenamefont {Liu},
  \citenamefont {Yuan}, \citenamefont {Zhang}, \citenamefont {Jin},
  \citenamefont {Narayan}, \citenamefont {Luo}, \citenamefont {Chen},
  \citenamefont {Yang}, \citenamefont {Zou}, \citenamefont {Wu}, \citenamefont
  {Sanvito}, \citenamefont {Xia}, \citenamefont {Li}, \citenamefont {Wang},\
  and\ \citenamefont {Xiu}}]{liu2016zeeman}%
  \BibitemOpen
  \bibfield  {author} {\bibinfo {author} {\bibfnamefont {Y.}~\bibnamefont
  {Liu}}, \bibinfo {author} {\bibfnamefont {X.}~\bibnamefont {Yuan}}, \bibinfo
  {author} {\bibfnamefont {C.}~\bibnamefont {Zhang}}, \bibinfo {author}
  {\bibfnamefont {Z.}~\bibnamefont {Jin}}, \bibinfo {author} {\bibfnamefont
  {A.}~\bibnamefont {Narayan}}, \bibinfo {author} {\bibfnamefont
  {C.}~\bibnamefont {Luo}}, \bibinfo {author} {\bibfnamefont {Z.}~\bibnamefont
  {Chen}}, \bibinfo {author} {\bibfnamefont {L.}~\bibnamefont {Yang}}, \bibinfo
  {author} {\bibfnamefont {J.}~\bibnamefont {Zou}}, \bibinfo {author}
  {\bibfnamefont {X.}~\bibnamefont {Wu}}, \bibinfo {author} {\bibfnamefont
  {S.}~\bibnamefont {Sanvito}}, \bibinfo {author} {\bibfnamefont
  {Z.}~\bibnamefont {Xia}}, \bibinfo {author} {\bibfnamefont {L.}~\bibnamefont
  {Li}}, \bibinfo {author} {\bibfnamefont {Z.}~\bibnamefont {Wang}}, \ and\
  \bibinfo {author} {\bibfnamefont {F.}~\bibnamefont {Xiu}},\ }\href
  {https://doi.org/10.1038/ncomms12516} {\bibfield  {journal} {\bibinfo
  {journal} {Nature Commun.}\ }\textbf {\bibinfo {volume} {7}},\ \bibinfo
  {pages} {12516} (\bibinfo {year} {2016})}\BibitemShut {NoStop}%
\bibitem [{\citenamefont {Chen}\ \emph
  {et~al.}(2015{\natexlab{a}})\citenamefont {Chen}, \citenamefont {Zhang},
  \citenamefont {Schneeloch}, \citenamefont {Zhang}, \citenamefont {Li},
  \citenamefont {Gu},\ and\ \citenamefont {Wang}}]{chen2015infrared}%
  \BibitemOpen
  \bibfield  {author} {\bibinfo {author} {\bibfnamefont {R.~Y.}\ \bibnamefont
  {Chen}}, \bibinfo {author} {\bibfnamefont {S.~J.}\ \bibnamefont {Zhang}},
  \bibinfo {author} {\bibfnamefont {J.~A.}\ \bibnamefont {Schneeloch}},
  \bibinfo {author} {\bibfnamefont {C.}~\bibnamefont {Zhang}}, \bibinfo
  {author} {\bibfnamefont {Q.}~\bibnamefont {Li}}, \bibinfo {author}
  {\bibfnamefont {G.~D.}\ \bibnamefont {Gu}}, \ and\ \bibinfo {author}
  {\bibfnamefont {N.~L.}\ \bibnamefont {Wang}},\ }\href {\doibase
  10.1103/PhysRevB.92.075107} {\bibfield  {journal} {\bibinfo  {journal} {Phys.
  Rev. B}\ }\textbf {\bibinfo {volume} {92}},\ \bibinfo {pages} {075107}
  (\bibinfo {year} {2015}{\natexlab{a}})}\BibitemShut {NoStop}%
\bibitem [{\citenamefont {Chen}\ \emph
  {et~al.}(2015{\natexlab{b}})\citenamefont {Chen}, \citenamefont {Chen},
  \citenamefont {Song}, \citenamefont {Schneeloch}, \citenamefont {Gu},
  \citenamefont {Wang},\ and\ \citenamefont {Wang}}]{chen2015magnetoinfrared}%
  \BibitemOpen
  \bibfield  {author} {\bibinfo {author} {\bibfnamefont {R.~Y.}\ \bibnamefont
  {Chen}}, \bibinfo {author} {\bibfnamefont {Z.~G.}\ \bibnamefont {Chen}},
  \bibinfo {author} {\bibfnamefont {X.-Y.}\ \bibnamefont {Song}}, \bibinfo
  {author} {\bibfnamefont {J.~A.}\ \bibnamefont {Schneeloch}}, \bibinfo
  {author} {\bibfnamefont {G.~D.}\ \bibnamefont {Gu}}, \bibinfo {author}
  {\bibfnamefont {F.}~\bibnamefont {Wang}}, \ and\ \bibinfo {author}
  {\bibfnamefont {N.~L.}\ \bibnamefont {Wang}},\ }\href
  {https://link.aps.org/doi/10.1103/PhysRevLett.115.176404} {\bibfield
  {journal} {\bibinfo  {journal} {Phys. Rev. Lett.}\ }\textbf {\bibinfo
  {volume} {115}},\ \bibinfo {pages} {176404} (\bibinfo {year}
  {2015}{\natexlab{b}})}\BibitemShut {NoStop}%
\bibitem [{\citenamefont {Xu}\ \emph {et~al.}(2018)\citenamefont {Xu},
  \citenamefont {Zhao}, \citenamefont {Marsik}, \citenamefont {Sheveleva},
  \citenamefont {Lyzwa}, \citenamefont {Dai}, \citenamefont {Chen},
  \citenamefont {Qiu},\ and\ \citenamefont {Bernhard}}]{xu2018temperature}%
  \BibitemOpen
  \bibfield  {author} {\bibinfo {author} {\bibfnamefont {B.}~\bibnamefont
  {Xu}}, \bibinfo {author} {\bibfnamefont {L.}~\bibnamefont {Zhao}}, \bibinfo
  {author} {\bibfnamefont {P.}~\bibnamefont {Marsik}}, \bibinfo {author}
  {\bibfnamefont {E.}~\bibnamefont {Sheveleva}}, \bibinfo {author}
  {\bibfnamefont {F.}~\bibnamefont {Lyzwa}}, \bibinfo {author} {\bibfnamefont
  {Y.}~\bibnamefont {Dai}}, \bibinfo {author} {\bibfnamefont {G.}~\bibnamefont
  {Chen}}, \bibinfo {author} {\bibfnamefont {X.}~\bibnamefont {Qiu}}, \ and\
  \bibinfo {author} {\bibfnamefont {C.}~\bibnamefont {Bernhard}},\ }\href
  {https://link.aps.org/doi/10.1103/PhysRevLett.121.187401} {\bibfield
  {journal} {\bibinfo  {journal} {Phys. Rev. Lett.}\ }\textbf {\bibinfo
  {volume} {121}},\ \bibinfo {pages} {187401} (\bibinfo {year}
  {2018})}\BibitemShut {NoStop}%
\bibitem [{\citenamefont {Yuan}\ \emph {et~al.}(2016)\citenamefont {Yuan},
  \citenamefont {Zhang}, \citenamefont {Liu}, \citenamefont {Narayan},
  \citenamefont {Song}, \citenamefont {Shen}, \citenamefont {Sui},
  \citenamefont {Xu}, \citenamefont {Yu}, \citenamefont {An}, \citenamefont
  {Zhao}, \citenamefont {Sanvito}, \citenamefont {Yan},\ and\ \citenamefont
  {Xiu}}]{yuan2016observation}%
  \BibitemOpen
  \bibfield  {author} {\bibinfo {author} {\bibfnamefont {X.}~\bibnamefont
  {Yuan}}, \bibinfo {author} {\bibfnamefont {C.}~\bibnamefont {Zhang}},
  \bibinfo {author} {\bibfnamefont {Y.}~\bibnamefont {Liu}}, \bibinfo {author}
  {\bibfnamefont {A.}~\bibnamefont {Narayan}}, \bibinfo {author} {\bibfnamefont
  {C.}~\bibnamefont {Song}}, \bibinfo {author} {\bibfnamefont {S.}~\bibnamefont
  {Shen}}, \bibinfo {author} {\bibfnamefont {X.}~\bibnamefont {Sui}}, \bibinfo
  {author} {\bibfnamefont {J.}~\bibnamefont {Xu}}, \bibinfo {author}
  {\bibfnamefont {H.}~\bibnamefont {Yu}}, \bibinfo {author} {\bibfnamefont
  {Z.}~\bibnamefont {An}}, \bibinfo {author} {\bibfnamefont {J.}~\bibnamefont
  {Zhao}}, \bibinfo {author} {\bibfnamefont {S.}~\bibnamefont {Sanvito}},
  \bibinfo {author} {\bibfnamefont {H.}~\bibnamefont {Yan}}, \ and\ \bibinfo
  {author} {\bibfnamefont {F.}~\bibnamefont {Xiu}},\ }\href
  {https://doi.org/10.1038/am.2016.166} {\bibfield  {journal} {\bibinfo
  {journal} {NPG Asia Materials}\ }\textbf {\bibinfo {volume} {8}},\ \bibinfo
  {pages} {e325} (\bibinfo {year} {2016})}\BibitemShut {NoStop}%
\bibitem [{\citenamefont {Chen}\ \emph {et~al.}(2017)\citenamefont {Chen},
  \citenamefont {Chen}, \citenamefont {Zhong}, \citenamefont {Schneeloch},
  \citenamefont {Zhang}, \citenamefont {Huang}, \citenamefont {Qu},
  \citenamefont {Yu}, \citenamefont {Li}, \citenamefont {Gu},\ and\
  \citenamefont {Wang}}]{chen2017spectroscopic}%
  \BibitemOpen
  \bibfield  {author} {\bibinfo {author} {\bibfnamefont {Z.-G.}\ \bibnamefont
  {Chen}}, \bibinfo {author} {\bibfnamefont {R.~Y.}\ \bibnamefont {Chen}},
  \bibinfo {author} {\bibfnamefont {R.~D.}\ \bibnamefont {Zhong}}, \bibinfo
  {author} {\bibfnamefont {J.}~\bibnamefont {Schneeloch}}, \bibinfo {author}
  {\bibfnamefont {C.}~\bibnamefont {Zhang}}, \bibinfo {author} {\bibfnamefont
  {Y.}~\bibnamefont {Huang}}, \bibinfo {author} {\bibfnamefont
  {F.}~\bibnamefont {Qu}}, \bibinfo {author} {\bibfnamefont {R.}~\bibnamefont
  {Yu}}, \bibinfo {author} {\bibfnamefont {Q.}~\bibnamefont {Li}}, \bibinfo
  {author} {\bibfnamefont {G.~D.}\ \bibnamefont {Gu}}, \ and\ \bibinfo {author}
  {\bibfnamefont {N.~L.}\ \bibnamefont {Wang}},\ }\href
  {https://www.pnas.org/content/114/5/816} {\bibfield  {journal} {\bibinfo
  {journal} {Proc. Natl. Acad. Sci. U.S.A.}\ }\textbf {\bibinfo {volume}
  {114}},\ \bibinfo {pages} {816} (\bibinfo {year} {2017})}\BibitemShut
  {NoStop}%
\bibitem [{\citenamefont {Pariari}\ and\ \citenamefont
  {Mandal}(2017)}]{pariari2017coexistence}%
  \BibitemOpen
  \bibfield  {author} {\bibinfo {author} {\bibfnamefont {A.}~\bibnamefont
  {Pariari}}\ and\ \bibinfo {author} {\bibfnamefont {P.}~\bibnamefont
  {Mandal}},\ }\href {https://doi.org/10.1038/srep40327} {\bibfield  {journal}
  {\bibinfo  {journal} {Sci. Rep.}\ }\textbf {\bibinfo {volume} {7}},\ \bibinfo
  {pages} {40327} (\bibinfo {year} {2017})}\BibitemShut {NoStop}%
\bibitem [{\citenamefont {Jiang}\ \emph {et~al.}(2017)\citenamefont {Jiang},
  \citenamefont {Dun}, \citenamefont {Zhou}, \citenamefont {Lu}, \citenamefont
  {Chen}, \citenamefont {Moon}, \citenamefont {Besara}, \citenamefont
  {Siegrist}, \citenamefont {Baumbach}, \citenamefont {Smirnov},\ and\
  \citenamefont {Jiang}}]{jiang2017landau}%
  \BibitemOpen
  \bibfield  {author} {\bibinfo {author} {\bibfnamefont {Y.}~\bibnamefont
  {Jiang}}, \bibinfo {author} {\bibfnamefont {Z.~L.}\ \bibnamefont {Dun}},
  \bibinfo {author} {\bibfnamefont {H.~D.}\ \bibnamefont {Zhou}}, \bibinfo
  {author} {\bibfnamefont {Z.}~\bibnamefont {Lu}}, \bibinfo {author}
  {\bibfnamefont {K.-W.}\ \bibnamefont {Chen}}, \bibinfo {author}
  {\bibfnamefont {S.}~\bibnamefont {Moon}}, \bibinfo {author} {\bibfnamefont
  {T.}~\bibnamefont {Besara}}, \bibinfo {author} {\bibfnamefont {T.~M.}\
  \bibnamefont {Siegrist}}, \bibinfo {author} {\bibfnamefont {R.~E.}\
  \bibnamefont {Baumbach}}, \bibinfo {author} {\bibfnamefont {D.}~\bibnamefont
  {Smirnov}}, \ and\ \bibinfo {author} {\bibfnamefont {Z.}~\bibnamefont
  {Jiang}},\ }\href {https://link.aps.org/doi/10.1103/PhysRevB.96.041101}
  {\bibfield  {journal} {\bibinfo  {journal} {Phys. Rev. B}\ }\textbf {\bibinfo
  {volume} {96}},\ \bibinfo {pages} {041101} (\bibinfo {year}
  {2017})}\BibitemShut {NoStop}%
\bibitem [{\citenamefont {Zhang}\ \emph
  {et~al.}(2019{\natexlab{b}})\citenamefont {Zhang}, \citenamefont {Wang},
  \citenamefont {Guo}, \citenamefont {Zhu}, \citenamefont {Yang}, \citenamefont
  {Wang}, \citenamefont {Qu}, \citenamefont {Pi}, \citenamefont {Lu},\ and\
  \citenamefont {Tian}}]{zhang2019thermoelectric}%
  \BibitemOpen
  \bibfield  {author} {\bibinfo {author} {\bibfnamefont {J.~L.}\ \bibnamefont
  {Zhang}}, \bibinfo {author} {\bibfnamefont {C.~M.}\ \bibnamefont {Wang}},
  \bibinfo {author} {\bibfnamefont {C.~Y.}\ \bibnamefont {Guo}}, \bibinfo
  {author} {\bibfnamefont {X.~D.}\ \bibnamefont {Zhu}}, \bibinfo {author}
  {\bibfnamefont {J.~Y.}\ \bibnamefont {Yang}}, \bibinfo {author}
  {\bibfnamefont {Y.~Q.}\ \bibnamefont {Wang}}, \bibinfo {author}
  {\bibfnamefont {Z.}~\bibnamefont {Qu}}, \bibinfo {author} {\bibfnamefont
  {L.}~\bibnamefont {Pi}}, \bibinfo {author} {\bibfnamefont {H.~Z.}\
  \bibnamefont {Lu}}, \ and\ \bibinfo {author} {\bibfnamefont {M.~L.}\
  \bibnamefont {Tian}},\ }\href {https://arxiv.org/abs/1904.00417v2} {\bibfield
   {journal} {\bibinfo  {journal} {arXiv:1904.00417}\ } (\bibinfo {year}
  {2019}{\natexlab{b}})}\BibitemShut {NoStop}%
\bibitem [{\citenamefont {Martino}\ \emph {et~al.}(2019)\citenamefont
  {Martino}, \citenamefont {Crassee}, \citenamefont {Eguchi}, \citenamefont
  {Santos-Cottin}, \citenamefont {Zhong}, \citenamefont {Gu}, \citenamefont
  {Berger}, \citenamefont {Rukelj}, \citenamefont {Orlita}, \citenamefont
  {Homes},\ and\ \citenamefont {Akrap}}]{martino2019two}%
  \BibitemOpen
  \bibfield  {author} {\bibinfo {author} {\bibfnamefont {E.}~\bibnamefont
  {Martino}}, \bibinfo {author} {\bibfnamefont {I.}~\bibnamefont {Crassee}},
  \bibinfo {author} {\bibfnamefont {G.}~\bibnamefont {Eguchi}}, \bibinfo
  {author} {\bibfnamefont {D.}~\bibnamefont {Santos-Cottin}}, \bibinfo {author}
  {\bibfnamefont {R.~D.}\ \bibnamefont {Zhong}}, \bibinfo {author}
  {\bibfnamefont {G.~D.}\ \bibnamefont {Gu}}, \bibinfo {author} {\bibfnamefont
  {H.}~\bibnamefont {Berger}}, \bibinfo {author} {\bibfnamefont
  {Z.}~\bibnamefont {Rukelj}}, \bibinfo {author} {\bibfnamefont
  {M.}~\bibnamefont {Orlita}}, \bibinfo {author} {\bibfnamefont {C.~C.}\
  \bibnamefont {Homes}}, \ and\ \bibinfo {author} {\bibfnamefont
  {A.}~\bibnamefont {Akrap}},\ }\href
  {https://link.aps.org/doi/10.1103/PhysRevLett.122.217402} {\bibfield
  {journal} {\bibinfo  {journal} {Phys. Rev. Lett.}\ }\textbf {\bibinfo
  {volume} {122}},\ \bibinfo {pages} {217402} (\bibinfo {year}
  {2019})}\BibitemShut {NoStop}%
\bibitem [{\citenamefont {Hohenberg}\ and\ \citenamefont
  {Kohn}(1964)}]{Hohenberg_Kohn}%
  \BibitemOpen
  \bibfield  {author} {\bibinfo {author} {\bibfnamefont {P.}~\bibnamefont
  {Hohenberg}}\ and\ \bibinfo {author} {\bibfnamefont {W.}~\bibnamefont
  {Kohn}},\ }\href {http://link.aps.org/doi/10.1103/PhysRev.136.B864}
  {\bibfield  {journal} {\bibinfo  {journal} {Phys. Rev.}\ }\textbf {\bibinfo
  {volume} {136}},\ \bibinfo {pages} {B864} (\bibinfo {year}
  {1964})}\BibitemShut {NoStop}%
\bibitem [{\citenamefont {Kohn}\ and\ \citenamefont {Sham}(1965)}]{Kohn_Sham}%
  \BibitemOpen
  \bibfield  {author} {\bibinfo {author} {\bibfnamefont {W.}~\bibnamefont
  {Kohn}}\ and\ \bibinfo {author} {\bibfnamefont {L.~J.}\ \bibnamefont
  {Sham}},\ }\href {http://link.aps.org/doi/10.1103/PhysRev.140.A1133}
  {\bibfield  {journal} {\bibinfo  {journal} {Phys. Rev.}\ }\textbf {\bibinfo
  {volume} {140}},\ \bibinfo {pages} {A1133} (\bibinfo {year}
  {1965})}\BibitemShut {NoStop}%
\bibitem [{\citenamefont {Kresse}\ and\ \citenamefont {Hafner}(1993)}]{vasp1}%
  \BibitemOpen
  \bibfield  {author} {\bibinfo {author} {\bibfnamefont {G.}~\bibnamefont
  {Kresse}}\ and\ \bibinfo {author} {\bibfnamefont {J.}~\bibnamefont
  {Hafner}},\ }\href {http://link.aps.org/doi/10.1103/PhysRevB.47.558}
  {\bibfield  {journal} {\bibinfo  {journal} {Phys. Rev. B}\ }\textbf {\bibinfo
  {volume} {47}},\ \bibinfo {pages} {558} (\bibinfo {year} {1993})}\BibitemShut
  {NoStop}%
\bibitem [{\citenamefont {Kresse}\ and\ \citenamefont {Hafner}(1994)}]{vasp2}%
  \BibitemOpen
  \bibfield  {author} {\bibinfo {author} {\bibfnamefont {G.}~\bibnamefont
  {Kresse}}\ and\ \bibinfo {author} {\bibfnamefont {J.}~\bibnamefont
  {Hafner}},\ }\href {http://link.aps.org/doi/10.1103/PhysRevB.49.14251}
  {\bibfield  {journal} {\bibinfo  {journal} {Phys. Rev. B}\ }\textbf {\bibinfo
  {volume} {49}},\ \bibinfo {pages} {14251} (\bibinfo {year}
  {1994})}\BibitemShut {NoStop}%
\bibitem [{\citenamefont {Kresse}\ and\ \citenamefont
  {Furthm\"uller}(1996{\natexlab{a}})}]{vasp3}%
  \BibitemOpen
  \bibfield  {author} {\bibinfo {author} {\bibfnamefont {G.}~\bibnamefont
  {Kresse}}\ and\ \bibinfo {author} {\bibfnamefont {J.}~\bibnamefont
  {Furthm\"uller}},\ }\href
  {http://www.sciencedirect.com/science/article/pii/0927025696000080}
  {\bibfield  {journal} {\bibinfo  {journal} {Comput. Mater. Sci.}\ }\textbf
  {\bibinfo {volume} {6}},\ \bibinfo {pages} {15} (\bibinfo {year}
  {1996}{\natexlab{a}})}\BibitemShut {NoStop}%
\bibitem [{\citenamefont {Kresse}\ and\ \citenamefont
  {Furthm\"uller}(1996{\natexlab{b}})}]{vasp4}%
  \BibitemOpen
  \bibfield  {author} {\bibinfo {author} {\bibfnamefont {G.}~\bibnamefont
  {Kresse}}\ and\ \bibinfo {author} {\bibfnamefont {J.}~\bibnamefont
  {Furthm\"uller}},\ }\href {http://link.aps.org/doi/10.1103/PhysRevB.54.11169}
  {\bibfield  {journal} {\bibinfo  {journal} {Phys. Rev. B}\ }\textbf {\bibinfo
  {volume} {54}},\ \bibinfo {pages} {11169} (\bibinfo {year}
  {1996}{\natexlab{b}})}\BibitemShut {NoStop}%
\bibitem [{\citenamefont {Bl\"ochl}(1994)}]{paw_original}%
  \BibitemOpen
  \bibfield  {author} {\bibinfo {author} {\bibfnamefont {P.~E.}\ \bibnamefont
  {Bl\"ochl}},\ }\href {http://link.aps.org/doi/10.1103/PhysRevB.50.17953}
  {\bibfield  {journal} {\bibinfo  {journal} {Phys. Rev. B}\ }\textbf {\bibinfo
  {volume} {50}},\ \bibinfo {pages} {17953} (\bibinfo {year}
  {1994})}\BibitemShut {NoStop}%
\bibitem [{\citenamefont {Kresse}\ and\ \citenamefont
  {Joubert}(1999)}]{paw_us_relation}%
  \BibitemOpen
  \bibfield  {author} {\bibinfo {author} {\bibfnamefont {G.}~\bibnamefont
  {Kresse}}\ and\ \bibinfo {author} {\bibfnamefont {D.}~\bibnamefont
  {Joubert}},\ }\href {http://link.aps.org/doi/10.1103/PhysRevB.59.1758}
  {\bibfield  {journal} {\bibinfo  {journal} {Phys. Rev. B}\ }\textbf {\bibinfo
  {volume} {59}},\ \bibinfo {pages} {1758} (\bibinfo {year}
  {1999})}\BibitemShut {NoStop}%
\bibitem [{\citenamefont {Monkhorst}\ and\ \citenamefont
  {Pack}(1976)}]{MP_grid}%
  \BibitemOpen
  \bibfield  {author} {\bibinfo {author} {\bibfnamefont {H.~J.}\ \bibnamefont
  {Monkhorst}}\ and\ \bibinfo {author} {\bibfnamefont {J.~D.}\ \bibnamefont
  {Pack}},\ }\href {http://link.aps.org/doi/10.1103/PhysRevB.13.5188}
  {\bibfield  {journal} {\bibinfo  {journal} {Phys. Rev. B}\ }\textbf {\bibinfo
  {volume} {13}},\ \bibinfo {pages} {5188} (\bibinfo {year}
  {1976})}\BibitemShut {NoStop}%
\bibitem [{\citenamefont {Perdew}\ \emph {et~al.}(1996)\citenamefont {Perdew},
  \citenamefont {Burke},\ and\ \citenamefont {Ernzerhof}}]{PBE}%
  \BibitemOpen
  \bibfield  {author} {\bibinfo {author} {\bibfnamefont {J.~P.}\ \bibnamefont
  {Perdew}}, \bibinfo {author} {\bibfnamefont {K.}~\bibnamefont {Burke}}, \
  and\ \bibinfo {author} {\bibfnamefont {M.}~\bibnamefont {Ernzerhof}},\ }\href
  {http://link.aps.org/doi/10.1103/PhysRevLett.77.3865} {\bibfield  {journal}
  {\bibinfo  {journal} {Phys. Rev. Lett.}\ }\textbf {\bibinfo {volume} {77}},\
  \bibinfo {pages} {3865} (\bibinfo {year} {1996})}\BibitemShut {NoStop}%
\bibitem [{\citenamefont {Paier}\ \emph
  {et~al.}(2006{\natexlab{a}})\citenamefont {Paier}, \citenamefont {Marsman},
  \citenamefont {Hummer}, \citenamefont {Kresse}, \citenamefont {Gerber},\ and\
  \citenamefont {\'{A}ngy\'{a}n}}]{hse06_functional}%
  \BibitemOpen
  \bibfield  {author} {\bibinfo {author} {\bibfnamefont {J.}~\bibnamefont
  {Paier}}, \bibinfo {author} {\bibfnamefont {M.}~\bibnamefont {Marsman}},
  \bibinfo {author} {\bibfnamefont {K.}~\bibnamefont {Hummer}}, \bibinfo
  {author} {\bibfnamefont {G.}~\bibnamefont {Kresse}}, \bibinfo {author}
  {\bibfnamefont {I.~C.}\ \bibnamefont {Gerber}}, \ and\ \bibinfo {author}
  {\bibfnamefont {J.~G.}\ \bibnamefont {\'{A}ngy\'{a}n}},\ }\href
  {http://scitation.aip.org/content/aip/journal/jcp/124/15/10.1063/1.2187006}
  {\bibfield  {journal} {\bibinfo  {journal} {J. Chem. Phys.}\ }\textbf
  {\bibinfo {volume} {124}},\ \bibinfo {pages} {154709} (\bibinfo {year}
  {2006}{\natexlab{a}})}\BibitemShut {NoStop}%
\bibitem [{\citenamefont {Paier}\ \emph
  {et~al.}(2006{\natexlab{b}})\citenamefont {Paier}, \citenamefont {Marsman},
  \citenamefont {Hummer}, \citenamefont {Kresse}, \citenamefont {Gerber},\ and\
  \citenamefont {\'{A}ngy\'{a}n}}]{hse06_functional_erratum}%
  \BibitemOpen
  \bibfield  {author} {\bibinfo {author} {\bibfnamefont {J.}~\bibnamefont
  {Paier}}, \bibinfo {author} {\bibfnamefont {M.}~\bibnamefont {Marsman}},
  \bibinfo {author} {\bibfnamefont {K.}~\bibnamefont {Hummer}}, \bibinfo
  {author} {\bibfnamefont {G.}~\bibnamefont {Kresse}}, \bibinfo {author}
  {\bibfnamefont {I.~C.}\ \bibnamefont {Gerber}}, \ and\ \bibinfo {author}
  {\bibfnamefont {J.~G.}\ \bibnamefont {\'{A}ngy\'{a}n}},\ }\href
  {http://scitation.aip.org/content/aip/journal/jcp/125/24/10.1063/1.2403866}
  {\bibfield  {journal} {\bibinfo  {journal} {J. Chem. Phys.}\ }\textbf
  {\bibinfo {volume} {125}},\ \bibinfo {pages} {249901} (\bibinfo {year}
  {2006}{\natexlab{b}})}\BibitemShut {NoStop}%
\bibitem [{\citenamefont {Koelling}\ and\ \citenamefont
  {Harmon}(1977)}]{soc_second_variation_method}%
  \BibitemOpen
  \bibfield  {author} {\bibinfo {author} {\bibfnamefont {D.~D.}\ \bibnamefont
  {Koelling}}\ and\ \bibinfo {author} {\bibfnamefont {B.~N.}\ \bibnamefont
  {Harmon}},\ }\href {http://stacks.iop.org/0022-3719/10/i=16/a=019} {\bibfield
   {journal} {\bibinfo  {journal} {J. Phys. C}\ }\textbf {\bibinfo {volume}
  {10}},\ \bibinfo {pages} {3107} (\bibinfo {year} {1977})}\BibitemShut
  {NoStop}%
\bibitem [{\citenamefont {Mostofi}\ \emph {et~al.}(2008)\citenamefont
  {Mostofi}, \citenamefont {Yates}, \citenamefont {Lee}, \citenamefont {Souza},
  \citenamefont {Vanderbilt},\ and\ \citenamefont
  {Marzari}}]{mostofi2008wannier90}%
  \BibitemOpen
  \bibfield  {author} {\bibinfo {author} {\bibfnamefont {A.~A.}\ \bibnamefont
  {Mostofi}}, \bibinfo {author} {\bibfnamefont {J.~R.}\ \bibnamefont {Yates}},
  \bibinfo {author} {\bibfnamefont {Y.-S.}\ \bibnamefont {Lee}}, \bibinfo
  {author} {\bibfnamefont {I.}~\bibnamefont {Souza}}, \bibinfo {author}
  {\bibfnamefont {D.}~\bibnamefont {Vanderbilt}}, \ and\ \bibinfo {author}
  {\bibfnamefont {N.}~\bibnamefont {Marzari}},\ }\href
  {https://www.sciencedirect.com/science/article/pii/S0010465507004936}
  {\bibfield  {journal} {\bibinfo  {journal} {Comput. Phys. Commun.}\ }\textbf
  {\bibinfo {volume} {178}},\ \bibinfo {pages} {685} (\bibinfo {year}
  {2008})}\BibitemShut {NoStop}%
\bibitem [{\citenamefont {Wu}\ \emph {et~al.}(2018)\citenamefont {Wu},
  \citenamefont {Zhang}, \citenamefont {Song}, \citenamefont {Troyer},\ and\
  \citenamefont {Soluyanov}}]{wu2018wanniertools}%
  \BibitemOpen
  \bibfield  {author} {\bibinfo {author} {\bibfnamefont {Q.}~\bibnamefont
  {Wu}}, \bibinfo {author} {\bibfnamefont {S.}~\bibnamefont {Zhang}}, \bibinfo
  {author} {\bibfnamefont {H.-F.}\ \bibnamefont {Song}}, \bibinfo {author}
  {\bibfnamefont {M.}~\bibnamefont {Troyer}}, \ and\ \bibinfo {author}
  {\bibfnamefont {A.~A.}\ \bibnamefont {Soluyanov}},\ }\href
  {https://www.sciencedirect.com/science/article/pii/S0010465517303442}
  {\bibfield  {journal} {\bibinfo  {journal} {Comput. Phys. Commun.}\ }\textbf
  {\bibinfo {volume} {224}},\ \bibinfo {pages} {405} (\bibinfo {year}
  {2018})}\BibitemShut {NoStop}%
\bibitem [{\citenamefont {Kunc}\ and\ \citenamefont
  {Martin}(1982)}]{phonon_finite_displacement}%
  \BibitemOpen
  \bibfield  {author} {\bibinfo {author} {\bibfnamefont {K.}~\bibnamefont
  {Kunc}}\ and\ \bibinfo {author} {\bibfnamefont {R.~M.}\ \bibnamefont
  {Martin}},\ }\href {http://link.aps.org/doi/10.1103/PhysRevLett.48.406}
  {\bibfield  {journal} {\bibinfo  {journal} {Phys. Rev. Lett.}\ }\textbf
  {\bibinfo {volume} {48}},\ \bibinfo {pages} {406} (\bibinfo {year}
  {1982})}\BibitemShut {NoStop}%
\bibitem [{\citenamefont {Lloyd-Williams}\ and\ \citenamefont
  {Monserrat}(2015)}]{non_diagonal}%
  \BibitemOpen
  \bibfield  {author} {\bibinfo {author} {\bibfnamefont {J.~H.}\ \bibnamefont
  {Lloyd-Williams}}\ and\ \bibinfo {author} {\bibfnamefont {B.}~\bibnamefont
  {Monserrat}},\ }\href {http://link.aps.org/doi/10.1103/PhysRevB.92.184301}
  {\bibfield  {journal} {\bibinfo  {journal} {Phys. Rev. B}\ }\textbf {\bibinfo
  {volume} {92}},\ \bibinfo {pages} {184301} (\bibinfo {year}
  {2015})}\BibitemShut {NoStop}%
\bibitem [{\citenamefont {Garate}(2013)}]{elph_topological_prl}%
  \BibitemOpen
  \bibfield  {author} {\bibinfo {author} {\bibfnamefont {I.}~\bibnamefont
  {Garate}},\ }\href {http://link.aps.org/doi/10.1103/PhysRevLett.110.046402}
  {\bibfield  {journal} {\bibinfo  {journal} {Phys. Rev. Lett.}\ }\textbf
  {\bibinfo {volume} {110}},\ \bibinfo {pages} {046402} (\bibinfo {year}
  {2013})}\BibitemShut {NoStop}%
\bibitem [{\citenamefont {Saha}\ and\ \citenamefont
  {Garate}(2014)}]{elph_topological_prb}%
  \BibitemOpen
  \bibfield  {author} {\bibinfo {author} {\bibfnamefont {K.}~\bibnamefont
  {Saha}}\ and\ \bibinfo {author} {\bibfnamefont {I.}~\bibnamefont {Garate}},\
  }\href {http://link.aps.org/doi/10.1103/PhysRevB.89.205103} {\bibfield
  {journal} {\bibinfo  {journal} {Phys. Rev. B}\ }\textbf {\bibinfo {volume}
  {89}},\ \bibinfo {pages} {205103} (\bibinfo {year} {2014})}\BibitemShut
  {NoStop}%
\bibitem [{\citenamefont {Kim}\ and\ \citenamefont
  {Jhi}(2015)}]{elph_topological_jhi}%
  \BibitemOpen
  \bibfield  {author} {\bibinfo {author} {\bibfnamefont {J.}~\bibnamefont
  {Kim}}\ and\ \bibinfo {author} {\bibfnamefont {S.-H.}\ \bibnamefont {Jhi}},\
  }\href {http://link.aps.org/doi/10.1103/PhysRevB.92.125142} {\bibfield
  {journal} {\bibinfo  {journal} {Phys. Rev. B}\ }\textbf {\bibinfo {volume}
  {92}},\ \bibinfo {pages} {125142} (\bibinfo {year} {2015})}\BibitemShut
  {NoStop}%
\bibitem [{\citenamefont {Monserrat}\ and\ \citenamefont
  {Vanderbilt}(2016)}]{monserrat_ti_temp}%
  \BibitemOpen
  \bibfield  {author} {\bibinfo {author} {\bibfnamefont {B.}~\bibnamefont
  {Monserrat}}\ and\ \bibinfo {author} {\bibfnamefont {D.}~\bibnamefont
  {Vanderbilt}},\ }\href
  {https://link.aps.org/doi/10.1103/PhysRevLett.117.226801} {\bibfield
  {journal} {\bibinfo  {journal} {Phys. Rev. Lett.}\ }\textbf {\bibinfo
  {volume} {117}},\ \bibinfo {pages} {226801} (\bibinfo {year}
  {2016})}\BibitemShut {NoStop}%
\bibitem [{\citenamefont {Antonius}\ and\ \citenamefont
  {Louie}(2016)}]{antonius_ti_temp}%
  \BibitemOpen
  \bibfield  {author} {\bibinfo {author} {\bibfnamefont {G.}~\bibnamefont
  {Antonius}}\ and\ \bibinfo {author} {\bibfnamefont {S.~G.}\ \bibnamefont
  {Louie}},\ }\href {https://link.aps.org/doi/10.1103/PhysRevLett.117.246401}
  {\bibfield  {journal} {\bibinfo  {journal} {Phys. Rev. Lett.}\ }\textbf
  {\bibinfo {volume} {117}},\ \bibinfo {pages} {246401} (\bibinfo {year}
  {2016})}\BibitemShut {NoStop}%
\bibitem [{\citenamefont {Wang}\ \emph {et~al.}(2017)\citenamefont {Wang},
  \citenamefont {Jo}, \citenamefont {Wu}, \citenamefont {Wu}, \citenamefont
  {Kaminski}, \citenamefont {Canfield},\ and\ \citenamefont
  {Johnson}}]{wang_typeII_elph}%
  \BibitemOpen
  \bibfield  {author} {\bibinfo {author} {\bibfnamefont {L.-L.}\ \bibnamefont
  {Wang}}, \bibinfo {author} {\bibfnamefont {N.~H.}\ \bibnamefont {Jo}},
  \bibinfo {author} {\bibfnamefont {Y.}~\bibnamefont {Wu}}, \bibinfo {author}
  {\bibfnamefont {Q.}~\bibnamefont {Wu}}, \bibinfo {author} {\bibfnamefont
  {A.}~\bibnamefont {Kaminski}}, \bibinfo {author} {\bibfnamefont {P.~C.}\
  \bibnamefont {Canfield}}, \ and\ \bibinfo {author} {\bibfnamefont {D.~D.}\
  \bibnamefont {Johnson}},\ }\href {\doibase 10.1103/PhysRevB.95.165114}
  {\bibfield  {journal} {\bibinfo  {journal} {Phys. Rev. B}\ }\textbf {\bibinfo
  {volume} {95}},\ \bibinfo {pages} {165114} (\bibinfo {year}
  {2017})}\BibitemShut {NoStop}%
\bibitem [{\citenamefont {Monserrat}\ and\ \citenamefont
  {Vanderbilt}(2017)}]{monserrat2017temperature}%
  \BibitemOpen
  \bibfield  {author} {\bibinfo {author} {\bibfnamefont {B.}~\bibnamefont
  {Monserrat}}\ and\ \bibinfo {author} {\bibfnamefont {D.}~\bibnamefont
  {Vanderbilt}},\ }\href
  {https://journals.aps.org/prmaterials/abstract/10.1103/PhysRevMaterials.1.054201}
  {\bibfield  {journal} {\bibinfo  {journal} {Physical Review Materials}\
  }\textbf {\bibinfo {volume} {1}},\ \bibinfo {pages} {054201} (\bibinfo {year}
  {2017})}\BibitemShut {NoStop}%
\bibitem [{\citenamefont {M\"{o}ller}\ \emph {et~al.}(2017)\citenamefont
  {M\"{o}ller}, \citenamefont {Sawatzky}, \citenamefont {Franz},\ and\
  \citenamefont {Berciu}}]{moller_typeII_elph}%
  \BibitemOpen
  \bibfield  {author} {\bibinfo {author} {\bibfnamefont {M.~M.}\ \bibnamefont
  {M\"{o}ller}}, \bibinfo {author} {\bibfnamefont {G.~A.}\ \bibnamefont
  {Sawatzky}}, \bibinfo {author} {\bibfnamefont {M.}~\bibnamefont {Franz}}, \
  and\ \bibinfo {author} {\bibfnamefont {M.}~\bibnamefont {Berciu}},\ }\href
  {https://doi.org/10.1038/s41467-017-02442-y} {\bibfield  {journal} {\bibinfo
  {journal} {Nature Commun.}\ }\textbf {\bibinfo {volume} {8}},\ \bibinfo
  {pages} {2267} (\bibinfo {year} {2017})}\BibitemShut {NoStop}%
\bibitem [{\citenamefont {Trushin}\ and\ \citenamefont
  {G\"orling}(2018)}]{topo_electronic_temp}%
  \BibitemOpen
  \bibfield  {author} {\bibinfo {author} {\bibfnamefont {E.}~\bibnamefont
  {Trushin}}\ and\ \bibinfo {author} {\bibfnamefont {A.}~\bibnamefont
  {G\"orling}},\ }\href {\doibase 10.1103/PhysRevLett.120.146401} {\bibfield
  {journal} {\bibinfo  {journal} {Phys. Rev. Lett.}\ }\textbf {\bibinfo
  {volume} {120}},\ \bibinfo {pages} {146401} (\bibinfo {year}
  {2018})}\BibitemShut {NoStop}%
\bibitem [{\citenamefont {Fjellv\r{a}g}\ and\ \citenamefont
  {Kjekshus}(1986)}]{zrte5_structure_Fjellvag_1986}%
  \BibitemOpen
  \bibfield  {author} {\bibinfo {author} {\bibfnamefont {H.}~\bibnamefont
  {Fjellv\r{a}g}}\ and\ \bibinfo {author} {\bibfnamefont {A.}~\bibnamefont
  {Kjekshus}},\ }\href
  {http://www.sciencedirect.com/science/article/pii/0038109886905363}
  {\bibfield  {journal} {\bibinfo  {journal} {Solid State Commun.}\ }\textbf
  {\bibinfo {volume} {60}},\ \bibinfo {pages} {91 } (\bibinfo {year}
  {1986})}\BibitemShut {NoStop}%
\bibitem [{\citenamefont {Fan}\ \emph {et~al.}(2017)\citenamefont {Fan},
  \citenamefont {Liang}, \citenamefont {Chen}, \citenamefont {Yao},\ and\
  \citenamefont {Zhou}}]{fan2017transition}%
  \BibitemOpen
  \bibfield  {author} {\bibinfo {author} {\bibfnamefont {Z.}~\bibnamefont
  {Fan}}, \bibinfo {author} {\bibfnamefont {Q.-F.}\ \bibnamefont {Liang}},
  \bibinfo {author} {\bibfnamefont {Y.}~\bibnamefont {Chen}}, \bibinfo {author}
  {\bibfnamefont {S.-H.}\ \bibnamefont {Yao}}, \ and\ \bibinfo {author}
  {\bibfnamefont {J.}~\bibnamefont {Zhou}},\ }\href
  {https://doi.org/10.1038/srep45667} {\bibfield  {journal} {\bibinfo
  {journal} {Sci. Rep.}\ }\textbf {\bibinfo {volume} {7}},\ \bibinfo {pages}
  {45667} (\bibinfo {year} {2017})}\BibitemShut {NoStop}%
\bibitem [{\citenamefont {Hinuma}\ \emph {et~al.}(2017)\citenamefont {Hinuma},
  \citenamefont {Pizzi}, \citenamefont {Kumagai}, \citenamefont {Oba},\ and\
  \citenamefont {Tanaka}}]{seek_path}%
  \BibitemOpen
  \bibfield  {author} {\bibinfo {author} {\bibfnamefont {Y.}~\bibnamefont
  {Hinuma}}, \bibinfo {author} {\bibfnamefont {G.}~\bibnamefont {Pizzi}},
  \bibinfo {author} {\bibfnamefont {Y.}~\bibnamefont {Kumagai}}, \bibinfo
  {author} {\bibfnamefont {F.}~\bibnamefont {Oba}}, \ and\ \bibinfo {author}
  {\bibfnamefont {I.}~\bibnamefont {Tanaka}},\ }\href
  {http://www.sciencedirect.com/science/article/pii/S0927025616305110}
  {\bibfield  {journal} {\bibinfo  {journal} {Comput. Mater. Sci.}\ }\textbf
  {\bibinfo {volume} {128}},\ \bibinfo {pages} {140 } (\bibinfo {year}
  {2017})}\BibitemShut {NoStop}%
\bibitem [{\citenamefont {Furuseth}\ \emph {et~al.}(1973)\citenamefont
  {Furuseth}, \citenamefont {Bratt\r{a}s},\ and\ \citenamefont
  {Kjekshus}}]{zrte5_structure_furuseth_1973}%
  \BibitemOpen
  \bibfield  {author} {\bibinfo {author} {\bibfnamefont {S.}~\bibnamefont
  {Furuseth}}, \bibinfo {author} {\bibfnamefont {L.}~\bibnamefont
  {Bratt\r{a}s}}, \ and\ \bibinfo {author} {\bibfnamefont {A.}~\bibnamefont
  {Kjekshus}},\ }\href
  {http://actachemscand.org/doi/10.3891/acta.chem.scand.27-2367} {\bibfield
  {journal} {\bibinfo  {journal} {Acta Chem. Scand.}\ }\textbf {\bibinfo
  {volume} {27}},\ \bibinfo {pages} {2367} (\bibinfo {year}
  {1973})}\BibitemShut {NoStop}%
\bibitem [{\citenamefont {Sodeck}\ \emph {et~al.}(1979)\citenamefont {Sodeck},
  \citenamefont {Mikler},\ and\ \citenamefont
  {Komarek}}]{zrte5_structure_Sodeck_1979}%
  \BibitemOpen
  \bibfield  {author} {\bibinfo {author} {\bibfnamefont {H.}~\bibnamefont
  {Sodeck}}, \bibinfo {author} {\bibfnamefont {H.}~\bibnamefont {Mikler}}, \
  and\ \bibinfo {author} {\bibfnamefont {K.~L.}\ \bibnamefont {Komarek}},\
  }\href {https://doi.org/10.1007/BF00903741} {\bibfield  {journal} {\bibinfo
  {journal} {Monatsh. Chem.}\ }\textbf {\bibinfo {volume} {110}},\ \bibinfo
  {pages} {1} (\bibinfo {year} {1979})}\BibitemShut {NoStop}%
\bibitem [{\citenamefont {Sambongi}\ \emph {et~al.}(1985)\citenamefont
  {Sambongi}, \citenamefont {Biljakovic}, \citenamefont {Smontara},\ and\
  \citenamefont {Guemas}}]{zrte5_structure_Sambongi_1985}%
  \BibitemOpen
  \bibfield  {author} {\bibinfo {author} {\bibfnamefont {T.}~\bibnamefont
  {Sambongi}}, \bibinfo {author} {\bibfnamefont {K.}~\bibnamefont
  {Biljakovic}}, \bibinfo {author} {\bibfnamefont {A.}~\bibnamefont
  {Smontara}}, \ and\ \bibinfo {author} {\bibfnamefont {L.}~\bibnamefont
  {Guemas}},\ }\href
  {http://www.sciencedirect.com/science/article/pii/0379677985901870}
  {\bibfield  {journal} {\bibinfo  {journal} {Synth. Met.}\ }\textbf {\bibinfo
  {volume} {10}},\ \bibinfo {pages} {161 } (\bibinfo {year}
  {1985})}\BibitemShut {NoStop}%
\bibitem [{\citenamefont {Hooton}(1955)}]{scha}%
  \BibitemOpen
  \bibfield  {author} {\bibinfo {author} {\bibfnamefont {D.~J.}\ \bibnamefont
  {Hooton}},\ }\href {https://doi.org/10.1080/14786440408520575} {\bibfield
  {journal} {\bibinfo  {journal} {Phil. Mag. Ser. 7}\ }\textbf {\bibinfo
  {volume} {46}},\ \bibinfo {pages} {422} (\bibinfo {year} {1955})}\BibitemShut
  {NoStop}%
\bibitem [{\citenamefont {He}\ \emph {et~al.}(2016)\citenamefont {He},
  \citenamefont {Thingna}, \citenamefont {Wang},\ and\ \citenamefont
  {Li}}]{scha_quartic}%
  \BibitemOpen
  \bibfield  {author} {\bibinfo {author} {\bibfnamefont {D.}~\bibnamefont
  {He}}, \bibinfo {author} {\bibfnamefont {J.}~\bibnamefont {Thingna}},
  \bibinfo {author} {\bibfnamefont {J.-S.}\ \bibnamefont {Wang}}, \ and\
  \bibinfo {author} {\bibfnamefont {B.}~\bibnamefont {Li}},\ }\href
  {https://link.aps.org/doi/10.1103/PhysRevB.94.155411} {\bibfield  {journal}
  {\bibinfo  {journal} {Phys. Rev. B}\ }\textbf {\bibinfo {volume} {94}},\
  \bibinfo {pages} {155411} (\bibinfo {year} {2016})}\BibitemShut {NoStop}%
\bibitem [{\citenamefont {Monserrat}(2016)}]{thermal_lines}%
  \BibitemOpen
  \bibfield  {author} {\bibinfo {author} {\bibfnamefont {B.}~\bibnamefont
  {Monserrat}},\ }\href {http://link.aps.org/doi/10.1103/PhysRevB.93.014302}
  {\bibfield  {journal} {\bibinfo  {journal} {Phys. Rev. B}\ }\textbf {\bibinfo
  {volume} {93}},\ \bibinfo {pages} {014302} (\bibinfo {year}
  {2016})}\BibitemShut {NoStop}%
\bibitem [{\citenamefont {Peng}\ \emph {et~al.}(2019)\citenamefont {Peng},
  \citenamefont {Bravic}, \citenamefont {MacManus-Driscoll},\ and\
  \citenamefont {Monserrat}}]{pbo2_temperature_topoloy}%
  \BibitemOpen
  \bibfield  {author} {\bibinfo {author} {\bibfnamefont {B.}~\bibnamefont
  {Peng}}, \bibinfo {author} {\bibfnamefont {I.}~\bibnamefont {Bravic}},
  \bibinfo {author} {\bibfnamefont {J.~L.}\ \bibnamefont {MacManus-Driscoll}},
  \ and\ \bibinfo {author} {\bibfnamefont {B.}~\bibnamefont {Monserrat}},\
  }\href {https://link.aps.org/doi/10.1103/PhysRevB.100.161101} {\bibfield
  {journal} {\bibinfo  {journal} {Phys. Rev. B}\ }\textbf {\bibinfo {volume}
  {100}},\ \bibinfo {pages} {161101} (\bibinfo {year} {2019})}\BibitemShut
  {NoStop}%
\bibitem [{\citenamefont {Manzoni}\ \emph {et~al.}(2015)\citenamefont
  {Manzoni}, \citenamefont {Sterzi}, \citenamefont {Crepaldi}, \citenamefont
  {Diego}, \citenamefont {Cilento}, \citenamefont {Zacchigna}, \citenamefont
  {Bugnon}, \citenamefont {Berger}, \citenamefont {Magrez}, \citenamefont
  {Grioni},\ and\ \citenamefont {Parmigiani}}]{manzoni2015ultrafast}%
  \BibitemOpen
  \bibfield  {author} {\bibinfo {author} {\bibfnamefont {G.}~\bibnamefont
  {Manzoni}}, \bibinfo {author} {\bibfnamefont {A.}~\bibnamefont {Sterzi}},
  \bibinfo {author} {\bibfnamefont {A.}~\bibnamefont {Crepaldi}}, \bibinfo
  {author} {\bibfnamefont {M.}~\bibnamefont {Diego}}, \bibinfo {author}
  {\bibfnamefont {F.}~\bibnamefont {Cilento}}, \bibinfo {author} {\bibfnamefont
  {M.}~\bibnamefont {Zacchigna}}, \bibinfo {author} {\bibfnamefont
  {P.}~\bibnamefont {Bugnon}}, \bibinfo {author} {\bibfnamefont
  {H.}~\bibnamefont {Berger}}, \bibinfo {author} {\bibfnamefont
  {A.}~\bibnamefont {Magrez}}, \bibinfo {author} {\bibfnamefont
  {M.}~\bibnamefont {Grioni}}, \ and\ \bibinfo {author} {\bibfnamefont
  {F.}~\bibnamefont {Parmigiani}},\ }\href
  {https://link.aps.org/doi/10.1103/PhysRevLett.115.207402} {\bibfield
  {journal} {\bibinfo  {journal} {Phys. Rev. Lett.}\ }\textbf {\bibinfo
  {volume} {115}},\ \bibinfo {pages} {207402} (\bibinfo {year}
  {2015})}\BibitemShut {NoStop}%
\bibitem [{\citenamefont {Skelton}\ \emph {et~al.}(1982)\citenamefont
  {Skelton}, \citenamefont {Wieting}, \citenamefont {Wolf}, \citenamefont
  {Fuller}, \citenamefont {Gubser}, \citenamefont {Francavilla},\ and\
  \citenamefont {Levy}}]{skelton1982giant}%
  \BibitemOpen
  \bibfield  {author} {\bibinfo {author} {\bibfnamefont {E.}~\bibnamefont
  {Skelton}}, \bibinfo {author} {\bibfnamefont {T.}~\bibnamefont {Wieting}},
  \bibinfo {author} {\bibfnamefont {S.}~\bibnamefont {Wolf}}, \bibinfo {author}
  {\bibfnamefont {W.}~\bibnamefont {Fuller}}, \bibinfo {author} {\bibfnamefont
  {D.~U.}\ \bibnamefont {Gubser}}, \bibinfo {author} {\bibfnamefont
  {T.}~\bibnamefont {Francavilla}}, \ and\ \bibinfo {author} {\bibfnamefont
  {F.}~\bibnamefont {Levy}},\ }\href
  {https://www.sciencedirect.com/science/article/pii/003810988291016X}
  {\bibfield  {journal} {\bibinfo  {journal} {Solid state communications}\
  }\textbf {\bibinfo {volume} {42}},\ \bibinfo {pages} {1} (\bibinfo {year}
  {1982})}\BibitemShut {NoStop}%
\bibitem [{\citenamefont {Nair}\ \emph {et~al.}(2018)\citenamefont {Nair},
  \citenamefont {Dumitrescu}, \citenamefont {Channa}, \citenamefont {Griffin},
  \citenamefont {Neaton}, \citenamefont {Potter},\ and\ \citenamefont
  {Analytis}}]{nair2018thermodynamic}%
  \BibitemOpen
  \bibfield  {author} {\bibinfo {author} {\bibfnamefont {N.~L.}\ \bibnamefont
  {Nair}}, \bibinfo {author} {\bibfnamefont {P.~T.}\ \bibnamefont
  {Dumitrescu}}, \bibinfo {author} {\bibfnamefont {S.}~\bibnamefont {Channa}},
  \bibinfo {author} {\bibfnamefont {S.~M.}\ \bibnamefont {Griffin}}, \bibinfo
  {author} {\bibfnamefont {J.~B.}\ \bibnamefont {Neaton}}, \bibinfo {author}
  {\bibfnamefont {A.~C.}\ \bibnamefont {Potter}}, \ and\ \bibinfo {author}
  {\bibfnamefont {J.~G.}\ \bibnamefont {Analytis}},\ }\href
  {https://link.aps.org/doi/10.1103/PhysRevB.97.041111} {\bibfield  {journal}
  {\bibinfo  {journal} {Phys. Rev. B}\ }\textbf {\bibinfo {volume} {97}},\
  \bibinfo {pages} {041111} (\bibinfo {year} {2018})}\BibitemShut {NoStop}%
\end{thebibliography}
\end{document}